\documentclass[12pt,preprint]{aastex}
\usepackage{amsmath}

\shortauthors{Hemsendorf, Sigurdsson, Spurzem}
\shorttitle{Binary Black Holes} 

\title{Collisional dynamics around binary black
holes in galactic centers} 

\author{Marc Hemsendorf\altaffilmark{1,3}, 
Steinn Sigurdsson\altaffilmark{2}, 
Rainer Spurzem\altaffilmark{3}}
\altaffiltext{1}{Department of Physics and Astronomy,Rutgers, the
State University of New Jersey, 136 Frelinghuysen Road, Piscataway, NJ
08854-8019, USA\\
e-mail: marchems@physics.rutgers.edu}
\altaffiltext{2}{Department of Astronomy and Astrophysics, 525 Davey
Lab, Penn State University, University Park, PA 16802, USA\\
e-mail: steinn@astro.psu.edu}
\altaffiltext{3}{Astronomisches Rechen-Institut, M\"onchhofstr. 12-14,
D-69120 Heidelberg, Germany\\
e-mail: spurzem@ari.uni-heidelberg.de}

\begin{document}

\begin{abstract}
We follow the sinking of two massive black holes in a spherical
stellar system where the black holes become bound under the influence
of dynamical friction. Once bound, the binary hardens by three-body
encounters with surrounding stars. We find that the binary wanders
inside the core, providing an enhanced supply of reaction partners for
the hardening. The binary evolves into a highly eccentric
orbit leading to coalescence well beyond a Hubble time. These are the
first results from a hybrid ``self consistent field'' (SCF) and direct
Aarseth $N$-body integrator (NBODY6), which combines the advantages of
the direct force calculation with the efficiency of the field
method. The code is designed for use on parallel architectures and is
therefore applicable to collisional $N$-body integrations with
extraordinarily large particle numbers ($> 10^5$). This creates the
possibility of simulating the dynamics of both globular clusters with
realistic collisional relaxation and stellar systems surrounding
supermassive black holes in galactic nuclei.
\end{abstract}

\section{Introduction}
Currently the standard picture of galaxy formation involves the
collapse of baryonic matter in hierarchically clustering dark matter
halos and the subsequent building of big galaxies from small ones via
merging processes e.g., \citep{Peebles:93, Diaferio:99, Kauffmann:99a,
Kauffmann:99b}.  While recent cosmological simulations can adequately
reproduce many global properties of galaxies and their correlations,
the details are still very much dependent on the gas physics and
stellar feedback involved (see e.g., \cite{Navarro:2000}).
Additionally, most, if not all, galaxies harbor supermassive black
holes in their center \citep{Magorrian:98, Richstone:98,
Kormendy:95}. Correlations have been recently detected between black
hole masses, galaxy masses, and central velocity dispersions in
galaxies \citep{Ferrarese:2000,Gebhardt:2000}. These correlations are
strong evidence that black holes in galactic nuclei are linked to the
dynamical history of their host galaxies. \citet{Haehnelt:2000} and
\citet{Kauffmann:2000} demonstrate how this is consistent with the
framework of semi-analytic models that follow the formation and
evolution of galaxies in a cold dark matter-dominated universe. They
assume supermassive black holes are formed and fueled during major
mergers, qualitatively explaining many aspects of the observed
evolution of galaxies, including the observed relation between bulge
luminosity, velocity dispersion, and central black hole mass. As
already discussed by \citet{Begelman:80}, such a scenario requires the
formation of galactic nuclei containing at least two black holes,
depending on the black hole merger rate relative to the galaxy merger
rate. However, there is very little observational evidence for massive
black hole binaries \citep{Lehto:96,Halpern:2000}. This conflict
between theory and observations has become known as the ``sinking
black hole problem''. As an alternative to minimally impacting stellar
dynamical processes, \citet{Gould:2000} and \citet{Armitage:2002} have
proposed mechanisms which lead to rapid decay of massive black hole
orbits and subsequent black hole mergers in galactic centers. Also,
\citet{Begelman:80} offered the solution that gas accretion could
dominate the orbital decay in the intermediate phase of the sinking
black hole problem when dynamical friction becomes
inefficient. However, as we will discuss later, dynamical friction, as
laid out by \citet{Chandrasekhar:43-a}, is not sufficiently effective
by itself to lead to rapid coalescence of black hole binaries.

If there are no quick mergers, multiple black hole nuclei could lose
black holes through slingshot ejections \citep{Valtonen:94}. Once a
binary system becomes hard, the high orbital velocities of the black
holes allow further hardening through close encounters and three-body
interactions with stars. Such processes will evacuate field stars from
the surroundings of the binary, therefore it can be argued that the stellar
scatterings cannot produce rapid coalescence. The preceding argument
assumes that the center of mass of the binary does not move with
respect to the stellar system. However, we will show that even with a
fairly symmetrical initial setup the binary gains some linear
momentum. This introduces a wandering motion which exceeds the
expectations from equipartition. The wandering of the binary
guarantees an adequate supply of stars for binary hardening and rapid
coalescence through purely stellar dynamical processes.

Our new computational method allows us to study in detail three-body
interactions of a black hole binary with field stars. Although one may
argue that the perturbing mass of the field stars is small compared to
the black hole mass and should have negligible impact, there are many
stars, and each encounter can lead to changes in binding energy and
eccentricity of the black hole binary. In fact, our models show that
the black hole binary keeps a rather high eccentricity due to the
encounters. Thus high eccentricity will speed up gravitational
radiation mergers very efficiently, and is, as noted by
\citet{Gould:2000} and \citet{Armitage:2002}, a way to expedite
massive black hole mergers in a purely stellar dynamical way.

The correct theoretical prediction of the frequency of black hole
mergers in galactic environments will be important in the search for
gravitational waves. The merging of supermassive black holes of
$3 \times 10^4$ to $3 \times 10^7 {\rm M}_\odot$ in the nuclei of merging
galaxies and protogalaxies can be detected with high signal-to-noise at
redshifts from $0 < z < 100$ \citep{Phinney:2000} by the Laser
Interferometer Space Antenna (LISA) \citep{Danzmann:2000}.

Previous attempts to quantify this prediction have been made by either
solving the perturbed two and three-body problem in simplified models
\citep{Mikkola:92}, direct $N$-body models
\citep{Makino:93,Makino:97}, or a combination of the two
\citep{Merritt:98-b,Quinlan:97}.  Simulating binary black hole
hardening is extremely challenging, algorithmically and
computationally. Since the mass differences between the black holes
and the stars is so large, high particle numbers are required in order
to model the relaxation processes around the black holes
accurately. The simulations have used softened particles on special
purpose computers \citep{Makino:93,Makino:97} or a hierarchical hybrid
code in which all forces involving the black hole particles are
Keplerian \citep{Merritt:98-b,Quinlan:97}. These schemes used particle
numbers in the order of $10^4$.

In this paper, we describe a new hybrid field-particle code which
treats all particles with orbits crossing the central regions of the
system with a high precision direct $N$-body method appropriate for
collisional stellar dynamics. All other particles are integrated using
a field method. In order to adapt both parts of the
hybrid code to each other, the field method (approximating the
potential exerted by a set of particles by a series expansion,
referred to here as ``SCF'') had to be upgraded to a fourth order
Hermite integrator. This integration also uses the time derivative of
the potential, as in modern direct $N$-body codes.

In the following sections some details of the sinking black hole
problem are introduced. Section \ref{sec:ES-intro} introduces the
integration software used for the numerical experiments described in
this paper. Section \ref{sec:ES-testing} is devoted to a comparison
between the new collisional code with a well used workhorse simulator
in this field called \emph{NBODY6} \citep{Aarseth:93,Aarseth:96,
Aarseth:99}, using its parallel implementation \emph{NBODY6++}
\citep{Spurzem:96, Spurzem:99-b}. The application of the code to the
sinking binary black hole problem is reported in section
\ref{sec:BlkH-runs}.

\section{Collisional stellar dynamics with {EuroStar}}
\label{sec:ES-intro}

\begin{figure}
\plotone{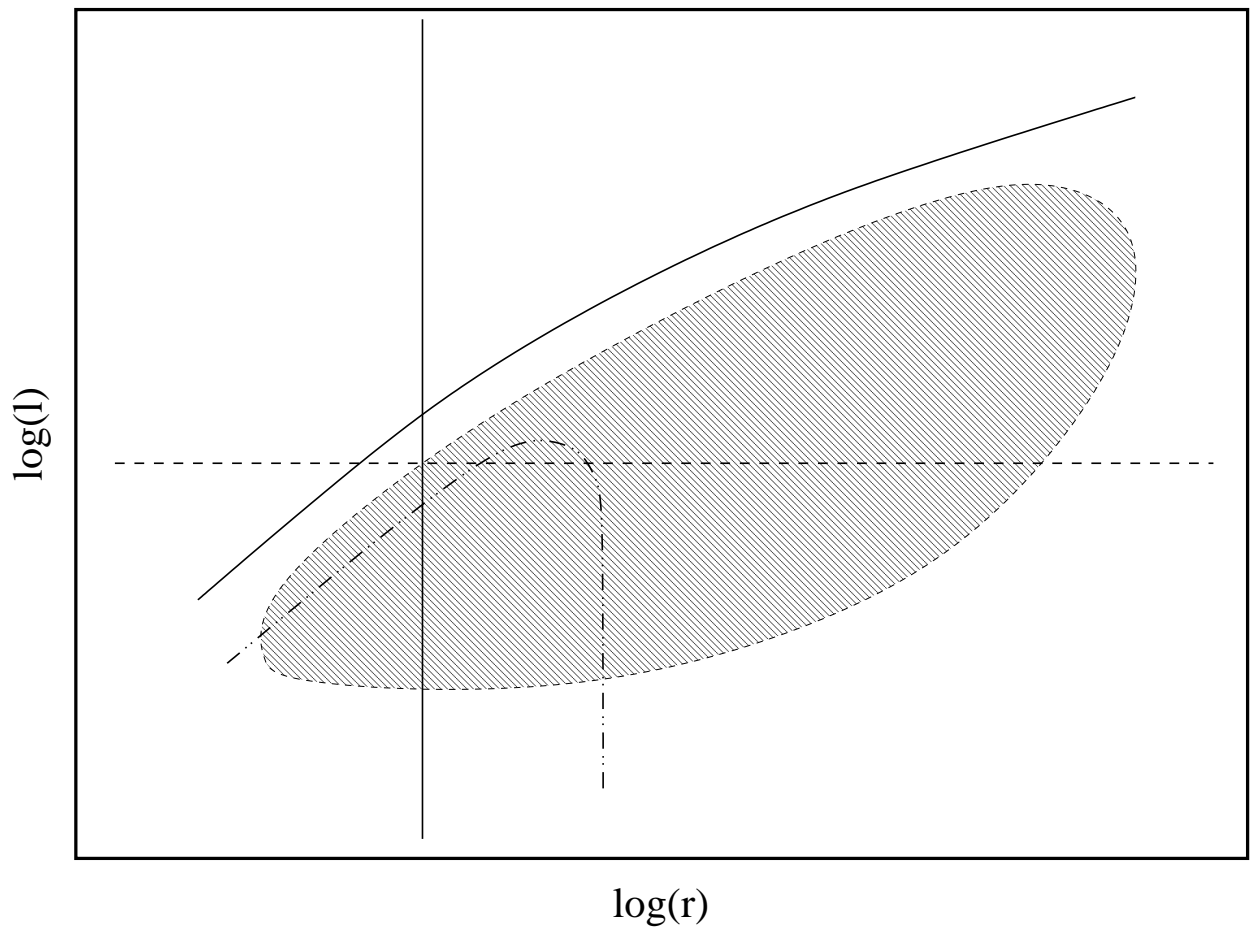}
\caption{Schematic decomposition of a star cluster according to its
angular momentum distribution. The hatched region symbolizes the
distribution of angular momentum per unit mass as a function of
radius. The solid curve above shows the angular momentum of particles
on circular orbits with escape velocity. As such, the solid curve
functions as an upper bound for the angular momentum at a given
radius, or as a lower limit for the radius a star can reach with fixed
angular momentum. If the cluster is divided into two parts by a
critical value for $l$, as shown by the horizontal dashed line, halo
particles which never reach the core can be distinguished from
particles which pass deep into the core (vertical solid
line). The dot-dot-dashed line shows the maximal angular momentum for
a set of particles selected by an energy criterion. Such a selection
would not affect all particles below that line. In fact, an energy
criterion is not sufficient for selecting all collisional particles in
our system.}
\label{fig:decomp-scheme}
\end{figure}

A numerical simulation of the hardening phase (until the massive black
holes start to radiate gravitational waves) must be able to accurately
follow three-body encounters. For this reason, the Keplerian potential
should not be softened in the denser parts of the
system. Computationally, the central part of the system would best be
treated using a collisional integrator. The code must be able to
integrate encounters leading to large angle deflections in an
efficient way, while requiring neither too much computing time nor
introducing energy errors. The overall $N$-body integration does not
need to be symplectic, but should keep the energy error as low as
possible.  On the other hand, in a system showing a core halo
structure, the bulk of the stars in the halo move under the influence
of the mean field of the whole cluster. The halo part of the central
galactic cluster can be integrated with a mean field method.

In the new method (which we will refer to as \textsc{EuroStar}), both
the collisional code {NBODY6++}
\citep{Aarseth:93,Aarseth:99,Spurzem:96} and the SCF method
\citep{Hernquist:92,Hernquist:95,Zhao:96,Sigurdsson:97,Holley:2001}
are merged to optimize large-$N$ collisional $N$-body simulations. The
star cluster, which is assumed to be in equilibrium, is divided into
two sections by applying a critical angular momentum criterion. As
shown in Figure \ref{fig:decomp-scheme}, this allows for distinction
between particles orbiting solely in the halo and ones which have
trajectories leading through the core of the system.

In a stationary gravitational point mass system, two-body relaxation
leads to an exchange between halo particles and core particles in a
divided cluster. In a system of more than $10^4$ particles, only a few
particles cross the core halo border per dynamical time. This is very
fortunate because it allows us to integrate the orbits of the halo
particles with a collisionless method and the core particles with a
collisional code. Exchanges of particles cause energy conservation
problems since the contribution of a particle to the main potential
would be changed from Keplerian to a sample point of a mean field in
{EuroStar}. This means that switching of particles from the core to
the halo part of the integrator and vice versa is not permitted.

NBODY6++ integrates trajectories of point masses in the
core of the system. It is an Aarseth-type direct force integrator
applying the Hermite integration scheme. NBODY6++ gains its efficiency
by implementing an Ahmad-Cohen neighbor scheme and individual block
time steps \citep{Ahmad:73,Aarseth:99}. Close interactions
between particles are treated by regularization of the equations of
motion \citep{Kustaanheimo:65}. NBODY6++ scales well on parallel
computer systems and can also be used with the GRAPE special purpose computer
\citep{Spurzem:99,Sugimoto:95}. The Hermite scheme
requires one to compute $\mathbf{F}_i$ and $\mathbf{\dot{F}}_i$ at each
time step where,
\begin{gather}
\mathbf{F}_i = - \sum_{j \ne i} \frac{G m_j \mathbf{r}_{ij}}{r^3_{ij}}, 
	\label{eq:DirektF}\\
\mathbf{\dot{F}}_i = - \sum_{j \ne i} G m_j \, \biggl[
	\frac{\mathbf{v}_{ij}}{r^3_{ij}} + \frac{3 \, (\mathbf{v}_{ij}
	\cdot \mathbf{r}_{ij} ) \, \mathbf{r}_{ij}}{r^5_{ij}}
	\biggr].
	\label{eq:DirektFdot}
\end{gather}
The relative distance between particles $i$ and $j$ is given by
$\mathbf{r}_{ij} = \mathbf{r}_{i} - \mathbf{r}_{j}$. Accordingly, the
relative velocity is $\mathbf{v}_{ij} = \mathbf{v}_{i} -
\mathbf{v}_{j}$.  The extra effort of computing two direct force
quantities allows one to approximate the particle's orbit to fourth
order. By storing $\mathbf{F}_i$ and $\mathbf{\dot{F}}_i$ from the
previous time step it is possible to interpolate the next two higher
derivatives and to apply a pre\-dic\-tor/cor\-rec\-tor scheme
\citep{Aarseth:96}.

The SCF method qualifies for the collisionless part of a spherical system
since then the basis functions are given analytically. This allows one to
implement the Hermite scheme for SCF, which makes SCF an ideal far
force extension to NBODY6++. Its drawback, however, is that this method
restricts the input systems to have an approximately spherical particle
distribution around the coordinate center. The possible asphericity
depends on the number of spherical harmonics used for the potential 
expansion. In order to have better convergence, one-parameter basis
functions for $\rho_{nlm}(\mathbf{r})$ and $\Phi_{nlm}(\mathbf{r})$
are used \citep{Zhao:96}:
\begin{gather}
\rho_{nlm}(\mathbf{r}) =  \sqrt{4 \pi} \, \frac{K_{nl} \, r^l \,
	C_n^{(\omega)}(\xi)}{r^{(2 - \frac{1}{\alpha})} \, (1 +
	r^{\frac{1}{\alpha}})^{2 + \alpha (2 l + 1)}} \, 
	Y_{lm}(\vartheta, \varphi), 
	\label{eq:SCFbasisRhoI} \\
\Phi_{nlm}(\mathbf{r}) = - \, \frac{\sqrt{4 \pi} \, r^l \, 
	C_n^{(\omega)}(\xi)}
	{(1 + r^{\frac{1}{\alpha}})^{\alpha(2 l + 1)}} \, 
	Y_{lm}(\vartheta, \varphi).
	\label{eq:SCFbasisPhiI}
\end{gather}
The $C_n^{(\omega)}$ are called ultraspherical or Gegenbauer
polynomials. The spherical harmonics are given by the
$Y_{lm}(\vartheta, \varphi)$. Once the $A_{nlm}$ for a certain set of
particles are known, an analytic expression of the potential and the
density is found. Due to the truncation of the expansion they
represent the mean field and mean density. This means that the force
at each position and its derivative can be computed using the
following expressions,
\begin{gather}
F(\mathbf{r}) = - \sum_{nlm} \, A_{nlm} \, \nabla
	\Phi_{nlm}(\mathbf{r}), 
	\label{eq:gradPhi}\\
\dot{F}(\mathbf{r}) = - \frac{d}{dt} \bigl( \, \sum_{nlm} \, A_{nlm}
	\, \nabla \Phi_{nlm}(\mathbf{r}) \bigr).
	\label{eq:dotgradPhi}
\end{gather}
Since Equations (\ref{eq:gradPhi}) and (\ref{eq:dotgradPhi}) lead to a
significant modification of the SCF scheme, we provide detailed form
of these expressions in the Appendix.

\section{Testing the hybrid code}
\label{sec:ES-testing}

As a first test for the new method, we have 
followed the last stages of an ongoing merger between two galaxies,
each containing a supermassive black hole. For the initial setup, it is
assumed that the stellar systems have already arranged themselves
into a spherical system. The two, formerly central, supermassive black
holes are moving through the stellar component with a speed on the
order of the relative velocity between the two initial galaxies.

In the present simulations, the stellar component is a realization of
a Plummer model. The density and potential of the spherically
symmetric Plummer model are given by \citep{Plummer:11}:
\begin{gather}
\rho(r) = \frac{3 M}{4 \pi} \; \frac{R^2}{(R^2 + r^2)^{5/2}}, \\
\Phi(r) = - G M \; \frac{1}{(R^2 + r^2)^{1/2}}.
\end{gather}
The quantity $M$ describes the total mass of the system and $G$ is the
gravitational constant.  With the Plummer radius chosen to be $R = 3
\pi/ 16$, the half mass radius of this system is at a radius of
$r_\mathrm{h} \approx 0.78$ in the model units of the simulations. The
total mass of the system $M$ is set to unity. The gravitational
constant $G$ is set to unity as well, conforming to the model units
described by Heggie \& Mathieu (1986). The stellar system is centered
around the origin. 

The black hole particles contain 1\% of the system's total mass. Their
initial positions are at $x = \pm0.5$, and their initial velocities
are 13.6\% of the circular velocity at their initial radii.

The black hole orbits are analyzed during the simulation assuming the
orbit can be approximated by the classical two-body problem. The
binding energies and eccentricities of the black hole orbits are
computed from their relative distances and velocities, assuming a
Keplerian potential. Once the black holes become bound, their two-body
attraction is the most important force. The eccentricity of the binary
$\epsilon$ and the binding energy $h$ are computed as follows, using
the definition $\mathbf{r} = \mathbf{r}_a - \mathbf{r}_b$. The vectors
$\mathbf{r}_a$ and $\mathbf{r}_b$ denote the position vectors of the
black holes $a$ and $b$. From this it follows that
\citep{Boccaletti:96},
\begin{gather}
h = \frac{1}{2} \, m_\mathrm{red}\, \dot{r}^2 +
	\frac{|\mathbf{l}|^2}{2 \, m_\mathrm{red} |\mathbf{r}|^2} -
	\frac{m_a \, m_b}{|\mathbf{r}|},
	\label{eq:hfull} \\
a = - \, \frac{m_a \, m_b}{2 \, h},
	\label{eq:afull} \\
\epsilon = \sqrt{1 + \frac{2 \, h \,
	|\mathbf{l}|^2}{m_\mathrm{red} \, (m_a \, m_b)^2}},
    \label{eq:efull}\\
\intertext{where $\mathbf{v} = \dot{\mathbf{r}}$, and }
\dot{r} = \frac{1}{|\mathbf{r}|} \, \mathbf{v} \cdot \mathbf{r}, \\
m_\mathrm{red} = \frac{m_a \, m_b}{m_a + m_b}, \\
\mathbf{l} = m_\mathrm{red} \, \mathbf{r} \times \mathbf{v}.
\label{eq:lfull}
\end{gather}
This method of analysis provides a sensitive measure for the moment
when the black holes become bound to one other. Furthermore, this way
of analyzing the data also offers a precise tool for following the
hardening of the binary.

\begin{figure}
\plotone{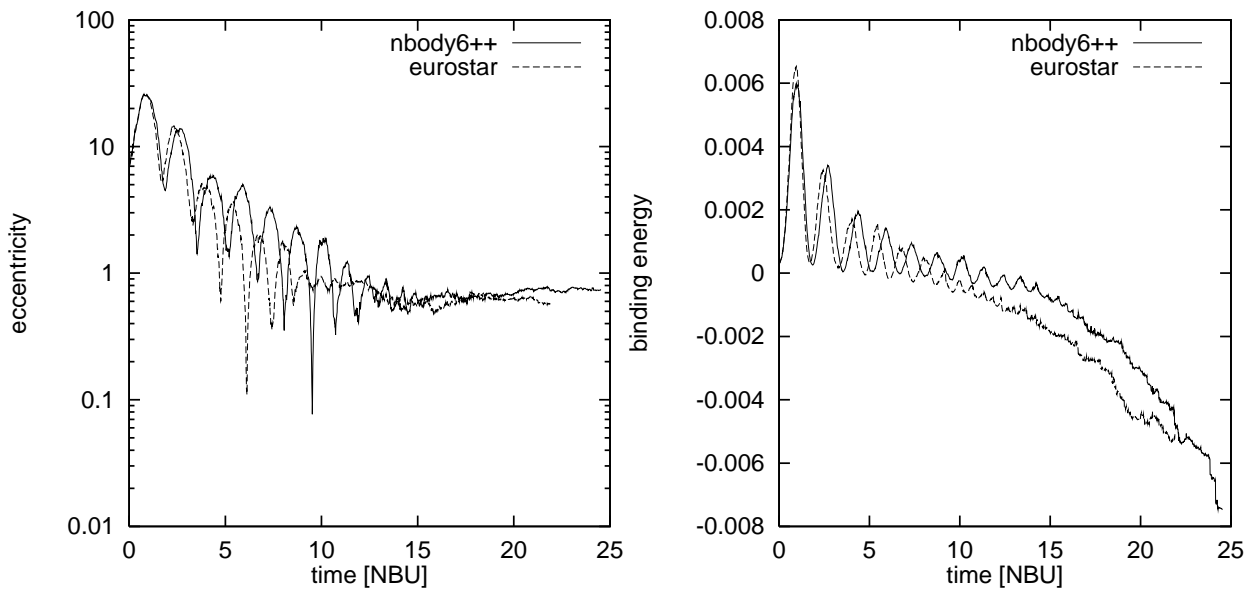}
\caption{Development of the orbital eccentricity of the black hole
binary as a function of time in $N$-body time units ($=$ NBU, left
graph) and its binding energy as a function of time (right). The
results for the direct method are shown by the solid line (NBODY6++),
the ones for the hybrid code (EuroStar) are given by the dashed
line. Both methods use the same initial model with 16384 particles.
Note, that if the binary is not yet bound, equations (\ref{eq:afull})
and (\ref{eq:efull}) formally yield values of $a>0$ and $e>1$. This
means that in that phase the black holes are still not yet
gravitationally bound to one other.}
\label{fig:esnb6bin-epshh}
\end{figure}

Since this sinking binary problem is the first application of
\textsc{EuroStar}, its results are compared with those of the fully
collisional code {NBODY6++}. For the comparison runs, 16384 particles
were simulated. Figure \ref{fig:esnb6bin-epshh} shows the
results for the two comparative runs.  The plot on the left hand side
in Figure \ref{fig:esnb6bin-epshh} shows the eccentricity of the
binary as a function of the simulated time in model units. The plot on
the right hand side shows the two-body binding energy as a function of
time. The binary becomes bound after 10 time units, in both cases.

The fully collisional method and the hybrid code {EuroStar} show
slightly different sinking rates for the binary at the beginning of
the simulation. These differences result from the different density of
collisional particles around the black holes in both codes. While the
black holes in the fully collisional run suffer small angle encounters
with every particle in the system, this is not possible in the hybrid
code. Naturally, all particles treated by the mean field method can
only interact with the system through changes in the mean
potential. After the binary has become bound, the hardening process is
driven by the stars which have a small enough impact parameter such
that they have an encounter timescale smaller than the orbital
timescale of the massive binary. Therefore, the hardening depends more
on the two-body encounters with neighboring particles, which have
large orbital velocities.

With increasing simulation time, the binary locks into an oscillating
motion around the center of mass of the stellar
component. This motion does not extend beyond of the
dense galactic core. Effectively, the differences in the density of
the collisional particles between the two methods vanish after the
binary has become bound. Figure \ref{fig:esnb6bin-epshh} reflects this
by showing a parallel evolution of eccentricity and binding energy for
times larger than 10 time units in the simulation.

\section{Hardening of a massive binary}
\label{sec:BlkH-runs}

This new code is intended for raising the total particle number for
collisional simulations of spherical $N$-body systems. 
Hence the evolution of one, two or several massive bodies in a dense
stellar cluster appears to be an ideal problem for EuroStar.
This is why we are addressing here the problem of a sinking
massive black hole binary in galactic centers. Another useful
potential application os the dynamics of globular clusters. 

\subsection{Initial conditions}
The particles representing the stellar component are distributed
according to Plummer's model with $R = 3 \pi/16$. The total mass of
the stars is fixed at $0.98$, while the black hole particles carry
$0.01$ each, so $M = 1.0$. This is a fairly high mass for the black
holes compared to the total mass of the stellar system, since
Ferrarese and Merritt \citet{Ferrarese:2000} found the black hole mass
in bulges to be smaller than that. However our simulations start at a
situation resembling the final stage of a galactic merger, which means
we are concentrating on the innermost part of the allready spherical
system.

The black hole particles are initially placed symmetrically about the
center of mass of the stellar component. Their initial radii are $r
\approx 0.64 r_\mathrm{h}$, their initial velocities are 13.6\% of the
circular velocity at this radius. In the given model units, this
represents starting points for the black holes at $x = \pm0.5$ and
$v_y = \pm0.1$. The center of mass of the stellar component is at the
origin. The mass factor between a stellar particle and a black hole
particle is: $1338.5$ for 131072 particles, $669.7$ for 65536
particles, and $335.4$ for 32768 particles.

In order to have a statistical basis for analysis, we compare
the results from five runs with 32768 particles, two runs with 65536
particles, and three runs with 131072 particles. Not all runs reached
the $60$ time unit mark due to time step scheduling problems caused by
accuracy problems in very close encounters between stars and a black
hole particle.  The regularization methods implemented in {EuroStar}
are identical to the ones suitable for open or globular cluster
simulations. The extreme situation in the late stages of the sinking
binary black hole problem may cause the chain algorithm to fail
\citep{Mikkola:93}. This problem can be solved by applying different
regularization methods.  However, up to the point of failure, the
simulations conserved the total energy with relative errors below
$10^{-4}$. In all runs, the binary becomes bound at approximately $10$
time units.

The parameters of the hybrid code have been adjusted in the following
way: The SCF part uses the parameter $\alpha = 0.5$ for the basis
functions. With this choice, the basis functions represent a Plummer
model to zeroth order, which is in accordance with the models used by
\citet{Clutton-Brock:73}. Also, this choice ensures an optimal
representation of the actual potential by the expansion method. To
allow flexibility in the expansion, seven basis functions are used
for the radial direction and five ($l=[0...5]$, $m=[-5...5]$) for the
angular expansions. The NBODY6++ part uses $\eta_i = 0.01$ for the
irregular time steps and $\eta_r = 0.02$ for the regular time
steps. Furthermore, the Ahmad-Cohen neighbor scheme \citep{Ahmad:73}
has been modified in such a way that the search radius for neighbors
is enhanced by a factor of $7.7$, $14.4$, and $27.7$ for interactions
with the black holes in the runs with
32768, 65536, and 131072 particles respectively.

\subsection{The motion of the massive bodies}

\begin{table}
\begin{tabular}{rrrrr}
$<t>_\mathrm{bin}$ & 32768 & 65536 & 131072 & all runs \\ \tableline
 2.5 &  2501 &  1000 &   2192 &   5693 \\
 7.5 &  2505 &  1508 &   3621 &   7634 \\
12.5 &  2681 &  4487 &  24536 &  31704 \\
17.5 &  3264 &  5581 &  32814 &  41659 \\
22.5 & 11987 &  6709 &  90249 & 108945 \\
27.5 &  3630 &  5528 &   7383 &  16541 \\
32.5 &  5766 &  2255 &   7568 &  15589 \\
37.5 &  4984 &  5541 &    254 &  10779 \\
47.5 &  1783 &  7447 &     -- &  10342 \\    
42.5 &  7768 &  2574 &     -- &   9230 \\
52.5 &  1440 &   637 &     -- &   2077 \\
57.5 &  1045 &  4567 &     -- &   5612 \\ \tableline
total  & 49354 & 47834 & 168617 & 265805
\end{tabular}
\caption{Number of sample points in the bins used for analyzing the
motion of the binary. The bins for the evolution time $t$ in $N$-body
units are centered around $<t>_\mathrm{bin}$. Plots using bins for the
total particle number $N_\mathrm{tot}$ have the number of sample
points given in the row labeled ``total''.}
\label{tab:samples}
\end{table}

In order to compare the runs, all data have been binned by the
parameter $t$, which represents the integrated time of the system in
$N$-body time units \citep{Heggie:86}. Table \ref{tab:samples} gives
the number of sample points for the orbital data of the black hole
particles. For technical reasons, the runs with 131072 particles
could not be continued to 60 time units. When 
binning the total particle number $N_\mathrm{tot}$, table
\ref{tab:samples} shows the number of samples in the row labeled
``total''. In the following, we present the results for the motion of
the black hole binary within the stellar system.

\subsubsection{Sinking rate of the binary}

\begin{figure}
\plottwo{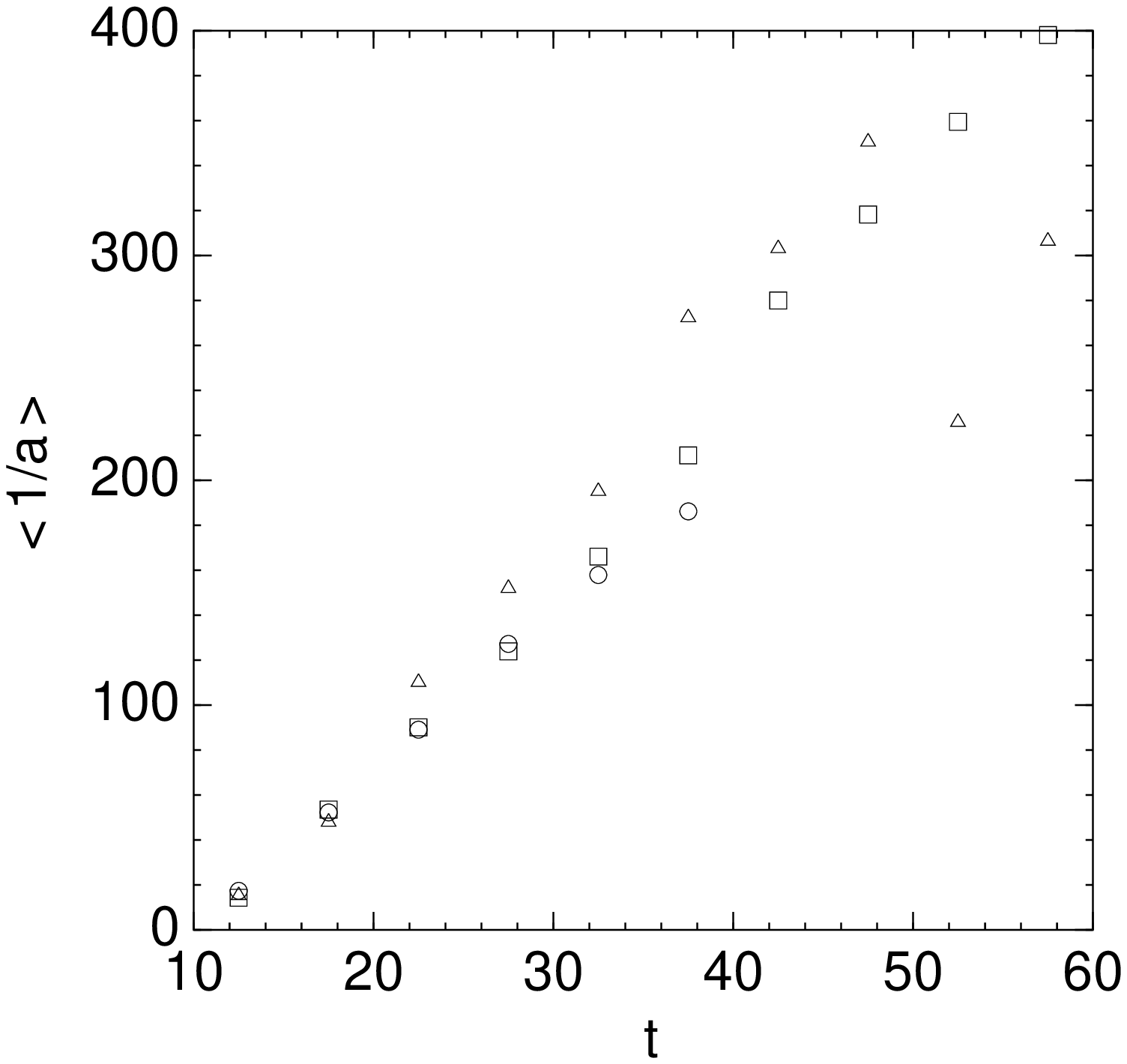}{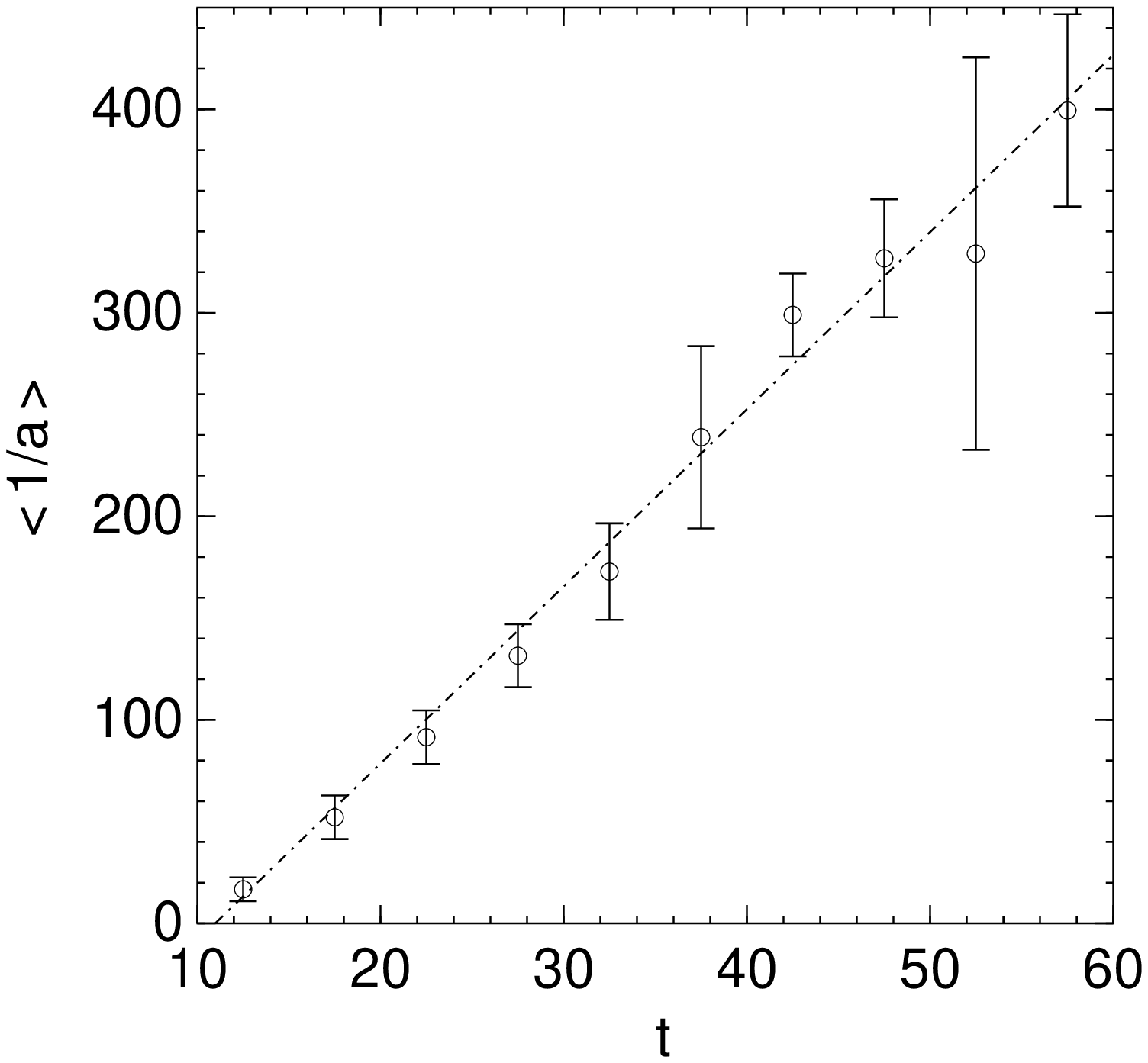}
\caption{Evolution of $\langle 1/a \rangle$ as a function of time,
after the binary becomes bound. The left plot shows the averages
computed for each particle group. 32768 data are plotted with open
triangles, 65536 data with open squares, and 131072 data with circles.
The averages over all runs carried out are shown in the right
plot. The error-bars indicate the standard deviation of the data in
the bins.}
\label{fig:oneatimall}
\end{figure}

Figure \ref{fig:oneatimall} shows the evolution of the quantity
$\langle 1/a \rangle$ as the average from the runs above. Equation
(\ref{eq:afull}) allows us to compute the semi-major axis $a$ from the
orbital data of the runs. The data is binned for comparison according
to the prescription above. From the data in each bin we evaluate the
average $\langle 1/a \rangle$ and the standard deviation. In order to
find the hardening rate, we fit a line to the averages, plotted as
circles in Figure \ref{fig:oneatimall}. The standard deviations of the
data points, given in the plot as the error-bars, supply the weighting
factors.

For the average over all runs plotted in Figure \ref{fig:oneatimall},
the regression line has a slope of $8.7 \pm 0.4$. The dependency on
the particle number can be deduced from the data shown on the left
side. The slopes are $9.6 \pm 0.5$ for the 32768 particle simulation,
$8.6 \pm 0.2$ for the 65536 particle simulation, and $6.8 \pm 0.2$ for
the 131072 particle simulation. There is clearly a dependence of the
results for the sinking rate on the particle number.  Compared with
other quantities we analyze in this study, the noise level in the data
for $1/a$ is low.  We observe strong interactions between the stellar
and the black hole particles in runs with 32768 particles. For this
reason, $1/a$ shows strong steplike changes in both directions. This
is most likely due to the small particle number.

\subsubsection{Evolution of the eccentricity}

\begin{figure}
\plotone{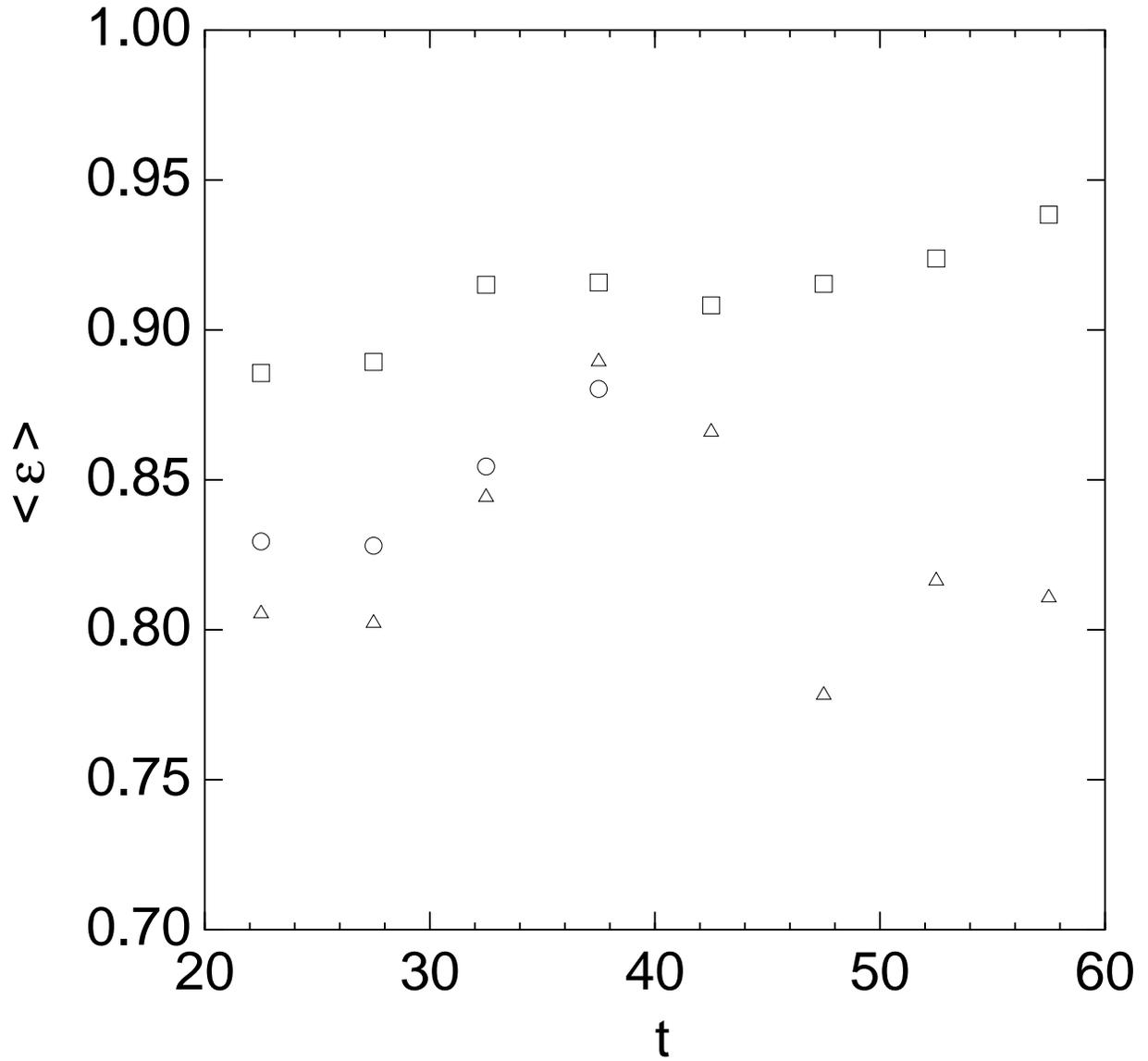}
\caption{Evolution of the eccentricity of the massive binary as a
  function of time. We are only plotting the data after the binary has
  become tightly bound in order to avoid unphysical values above $1$
  and strong scattering of the data. 32768 data are plotted with open
  triangles, 65536 data with open squares, and 131072 data with
  circles.}
\label{fig:eccentricity}
\end{figure}

We are only studying the evolution of the eccentricity after the
binary became bound. Because after 20 time units the eccentricity evolves
relatively smoothly for each run, we are concentrating our analysis on
the time range between 20 and 60 time units.  

Figure \ref{fig:eccentricity} shows the mean eccentricity binned in
time slots with a width of five time units. The symbols represent the
averages in these bins. 32768 data are plotted with open triangles,
65536 data with open squares, and 131072 data with circles. While the
eccentricities settle at values between $\epsilon = 0.6$ and $\epsilon
= 0.9$ for the runs with 32768 particles, the runs with higher
particle numbers show a fairly parallel evolution. The averages of the
32768 particle runs agree very much with the averages from 131072
data. However, the averages for the 65536 data are clearly higher. 

With our initial conditions, the binary evolves in a highly eccentric
orbit, which is around $\epsilon = 0.85$. The system retains this high
eccentricity until the end of our simulations. 

\subsubsection{Evolution of the angular momentum}

\begin{figure}
%\includegraphics[width=\linewidth]{Figures/theta-tim-all.eps}
%\plottwo{f19.eps}{f4.eps}
\plottwo{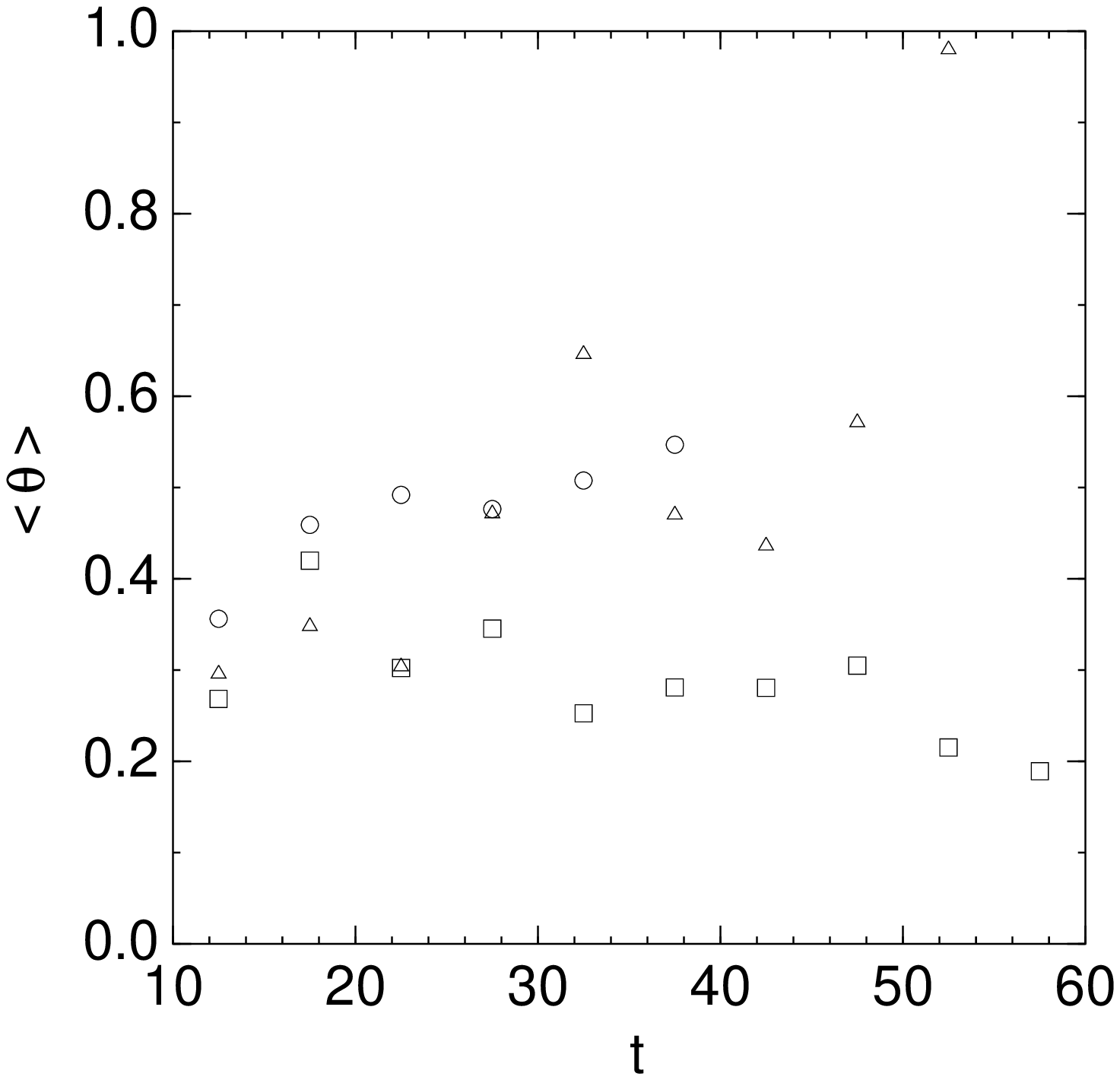}{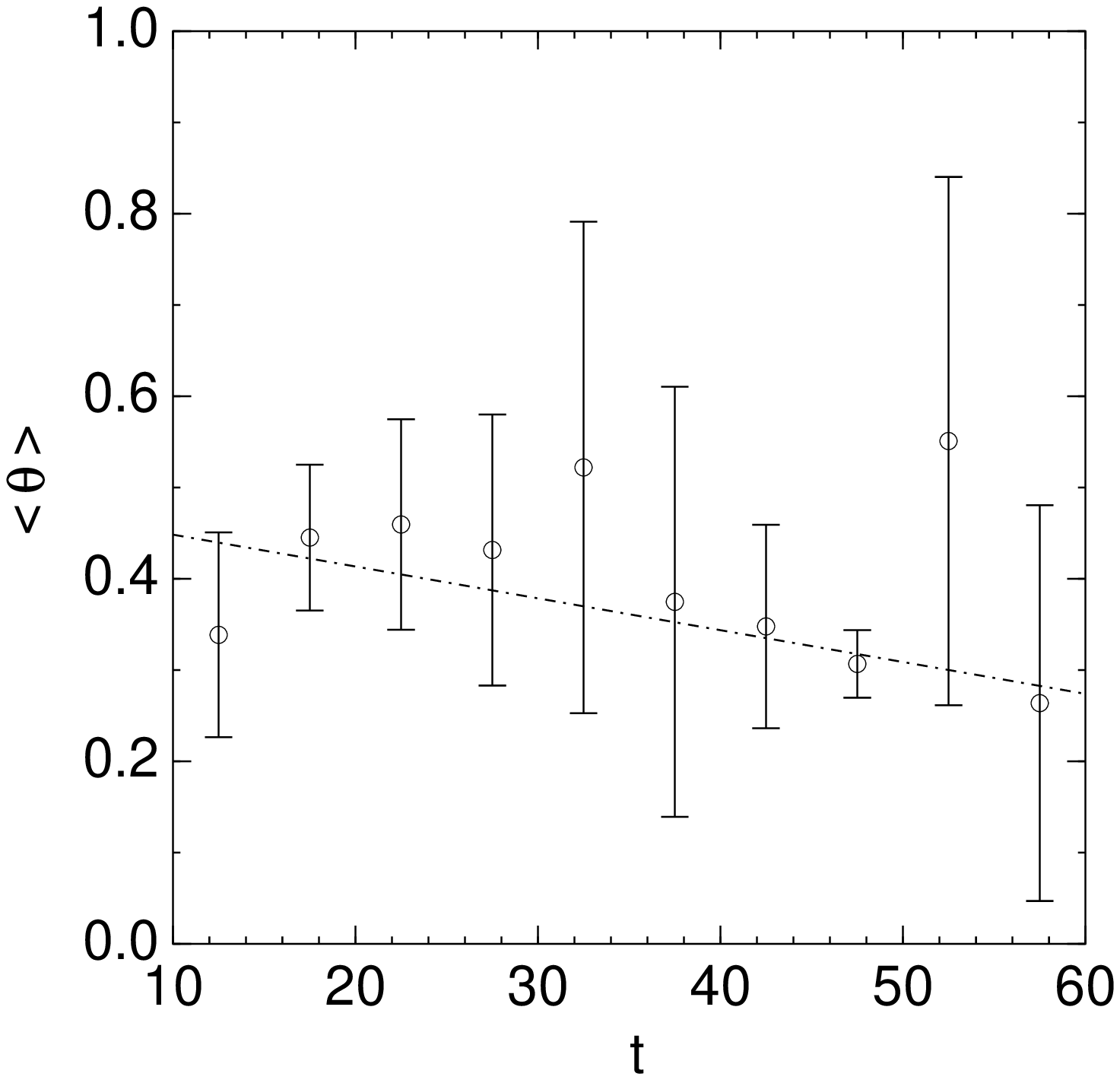}
\caption{Evolution of polar angle $\theta$ of the angular momentum of
the massive binary after it has become bound. The angular momentum
vector is initially aligned to the $z$-axis. The left plot shows
results for the particle groups: 32768 data are plotted with open
triangles, 65536 data with open squares, and 131072 data with
circles. The average over all data is shown on the right side.}
\label{fig:thetatimall}
\end{figure}

In order to study the evolution of the angular momentum of the bound
binary we plot the angle $\theta$ between the $z$-coordinate axis and
$\mathbf{l}/l$ versus time units in Figure \ref{fig:thetatimall}. 
$\theta$ is zero initially. As with $\langle 1/a \rangle$, all data from the
simulations are binned and averaged. The open circles in Figure
\ref{fig:thetatimall} represent the averages, while the error bars are the
standard deviations in the data. 

Once the binary becomes bound, the $\theta$ changes only
slightly in all simulations. Averaged over the time between the
first bound orbit of the massive particles and the end of the
simulations, the average value of $\theta$ becomes $0.5 \pm 0.3$ for the
32768 runs, $0.3 \pm 0.1$ for the 65536 runs, and $0.5 \pm 0.1$ for
the 131072 runs. 

The results in Figure \ref{fig:thetatimall} can be fitted by a
straight line. The slope of this line is $-0.003 \pm 0.003$. When we
group the simulations according to particle number, the fitting lines
have slopes of $0.016 \pm 0.004$ for the runs with 32768 particles,
$-0.0044 \pm 0.0006$ for the 65536 runs, and $0.006 \pm 0.002$ for the
131072 runs. Though small, these slopes are all significantly nonzero
and distinct from one other. Torques clearly act on the binary system
throughout the simulations. The data for the runs with 32768 particles
and with the small mass ratio between black holes and stellar
particles is very noisy and shows steplike changes in $\theta$.

While $\theta$ evolves in an ordered way until the binary becomes
bound, the angle $\phi$ between the $x$-coordinate axis and the
normalized angular momentum vector behaves more randomly. Until the
massive particles become bound, $\phi$ changes rapidly reaching all
values between $0$ and $2\pi$. However, once the binary becomes bound,
$\phi$ settles to a single value for each run. The changes in $\phi$
are subsequently of the same order of magnitude as for $\theta$.

\subsubsection{Wandering motion of the binary}

\begin{figure}
%\includegraphics[width=\linewidth]{Figures/barrsq-cmd-tim-all.eps}
%\plotone{f5.eps}
%\plottwo{f13.eps}{f5.eps}
\plottwo{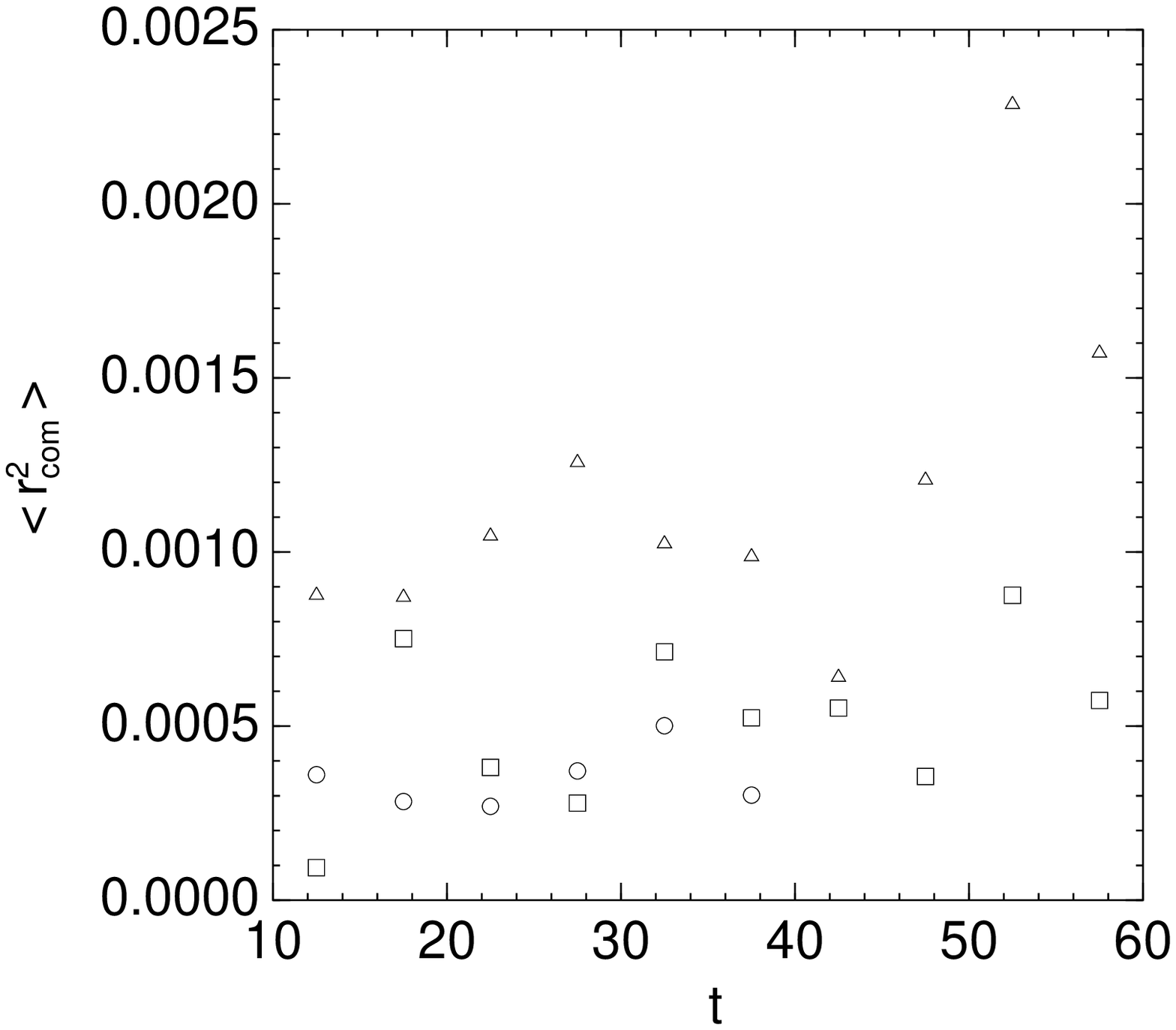}{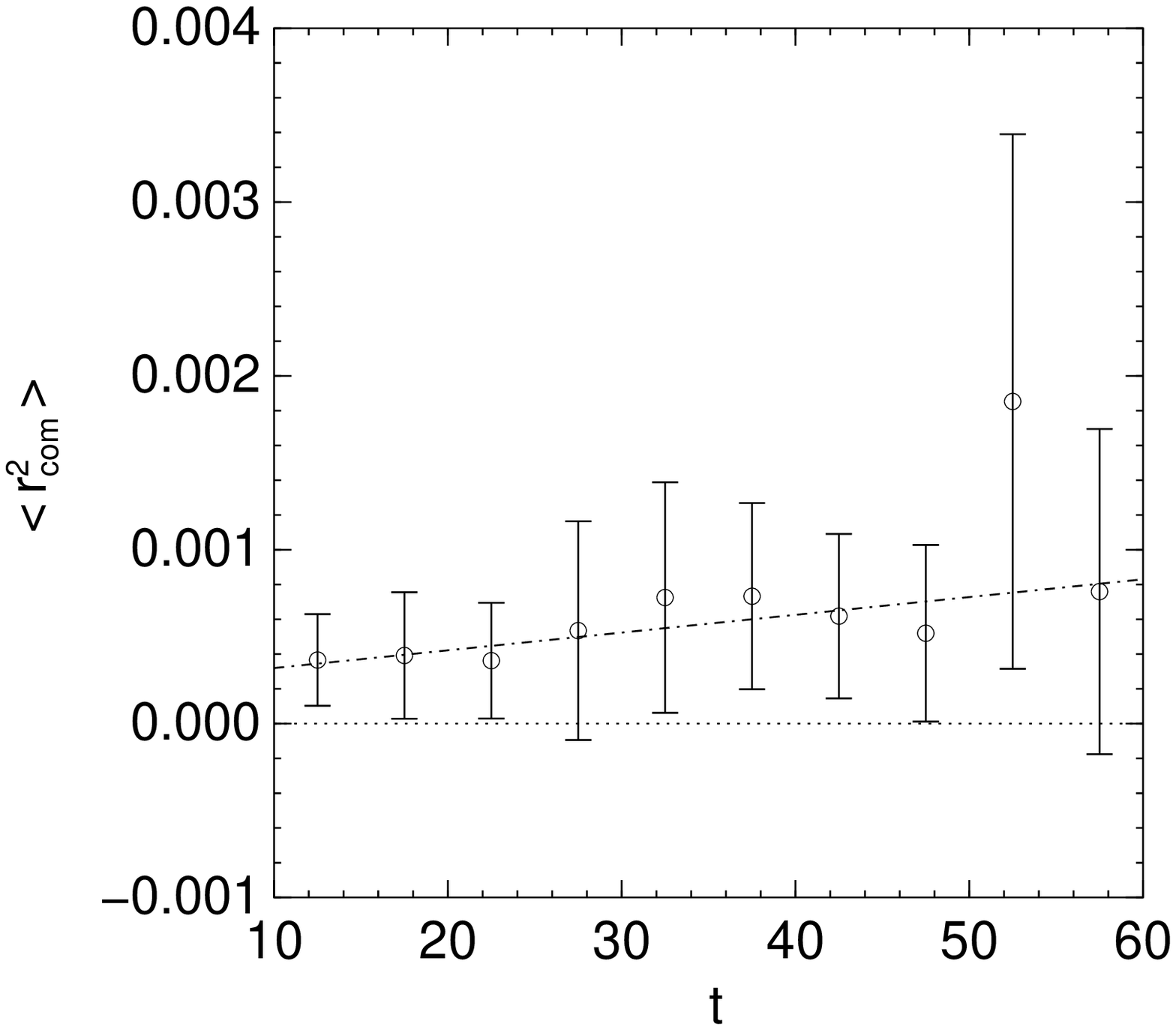}
\caption{The mean of $r^2_\mathrm{com}$ taken over the all simulations
as a function of the integrated time in $N$-body units. The quantity
$r_\mathrm{com}$ is the distance of the center of mass of the black
hole binary to the center of mass of the stellar particles. The left
plot shows results for the particle groups: 32768 data are plotted
with open triangles, 65536 data with open squares, and 131072 data
with circles. The average over all data is shown on the right side.}
\label{fig:barrsqcomtime}
\end{figure}

\begin{figure}
\plotone{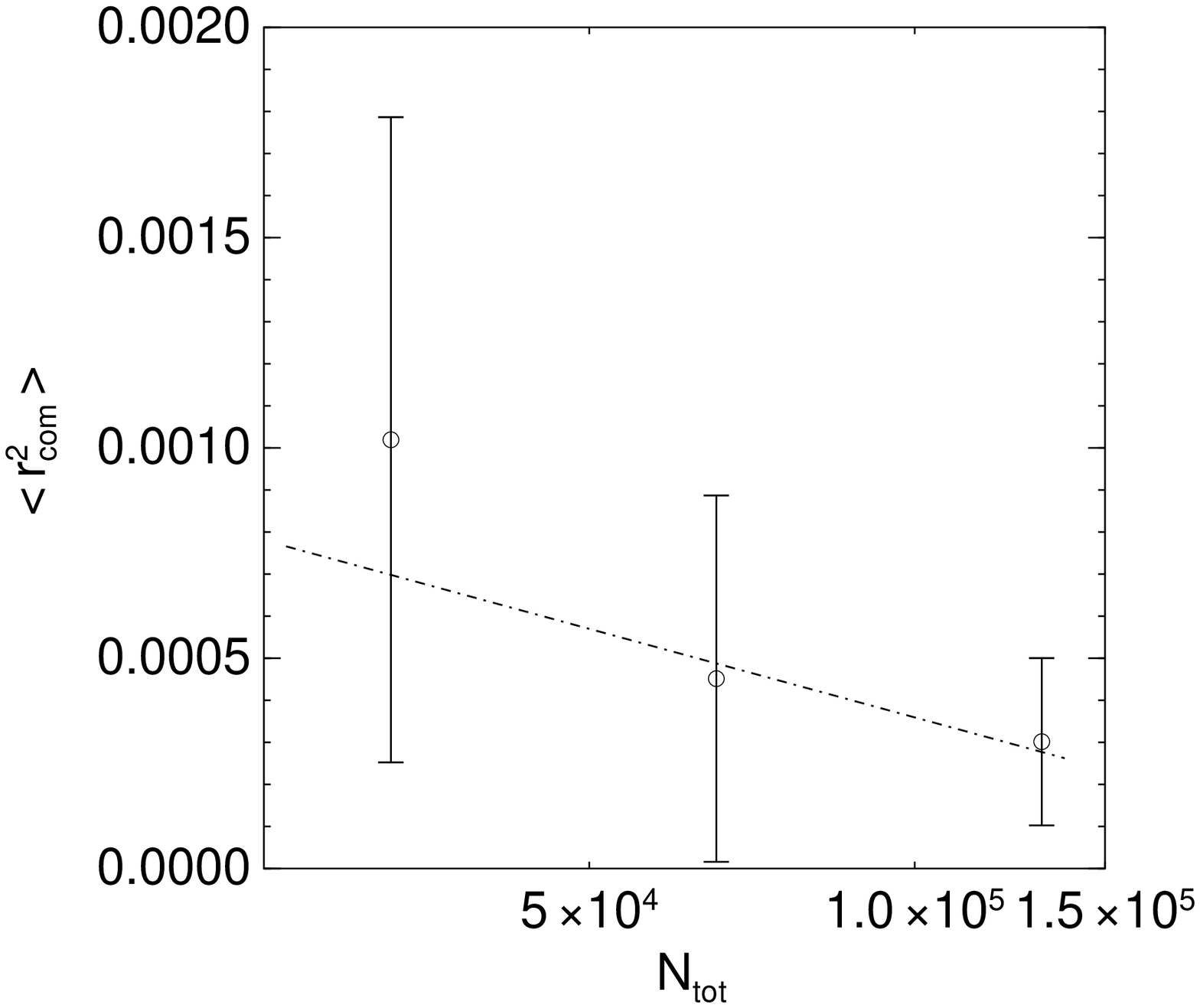}
\caption{The mean of $r^2_\mathrm{com}$ taken over the full integrated
times as a function of the total number of particles in the
simulations.  The quantity $r_\mathrm{com}$ is defined as in Figure
\ref{fig:barrsqcomtime}.}
\label{fig:barrsqcomntot}
\end{figure}

Studying the wandering of the binary using the quantity $\langle
r^2_\mathrm{com} \rangle$, we can compare the observed motion to the
expected Brownian motion in the system. $r_\mathrm{com}$ is the
distance from the center of mass of the black hole binary to the
center of mass of the stellar system.  Figure \ref{fig:barrsqcomtime}
implies that the mean motion is not constant with time. However, since
the slope of the fitting line is $(1.0 \pm 1.1) \times 10^{-5}$, the
behavior is constant within $1\sigma$ uncertainty. For the individual
particle number groups the situation is as follows: For 32768
particles we find a slope of $(0.6 \pm 1.5) \times 10^{-5}$, for 65536
particles a slope of $(1.0 \pm 0.7) \times 10^{-5}$, and for 131072
particles a slope of $(0.9 \pm 7.8) \times 10^{-6}$.  Compared to its
mean value over the whole simulation, the evolution of $\langle
r^2_\mathrm{com} \rangle$ with time introduces changes of not more
than 10\%. For this reason, we assume $\langle r^2_\mathrm{com}
\rangle$ to be constant for the analysis of the Brownian motion.
Figure \ref{fig:barrsqcomntot} shows the mean squared distance between
the center of mass of the black hole system and the stellar system as
a function of the total particle number of the simulations.  The slope
of the fitting line is $(-4.5 \pm 5.6) \times 10^{-9}$. Given our small
sample of runs we cannot determine a dependency of $\langle
r^2_\mathrm{com} \rangle$ on the particle number.

\begin{figure}
%\includegraphics[width=\linewidth]{Figures/barvsq-cmd-tim-all.eps}
%\plotone{f8.eps}
%\plottwo{f14.eps}{f8.eps}
\plottwo{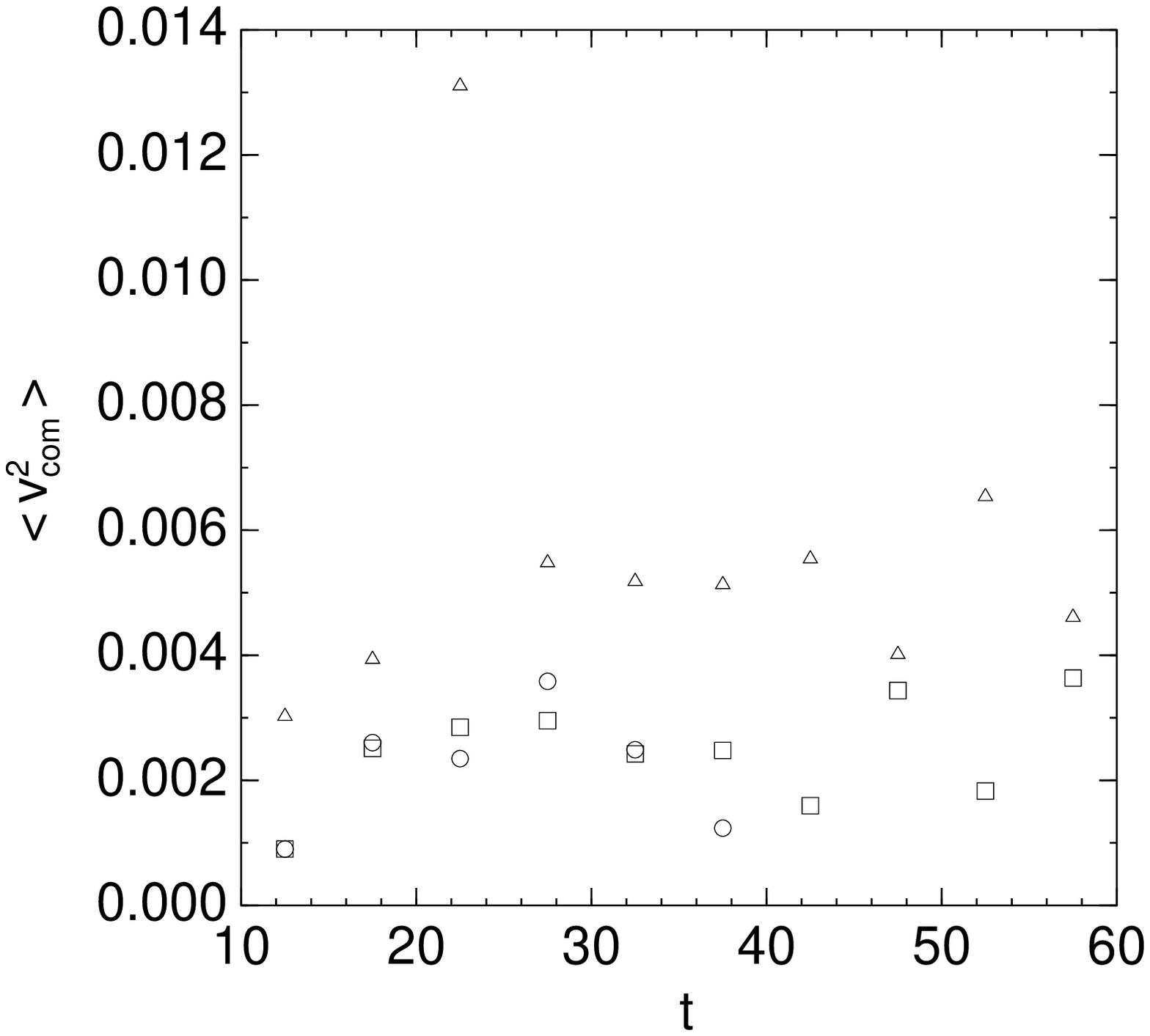}{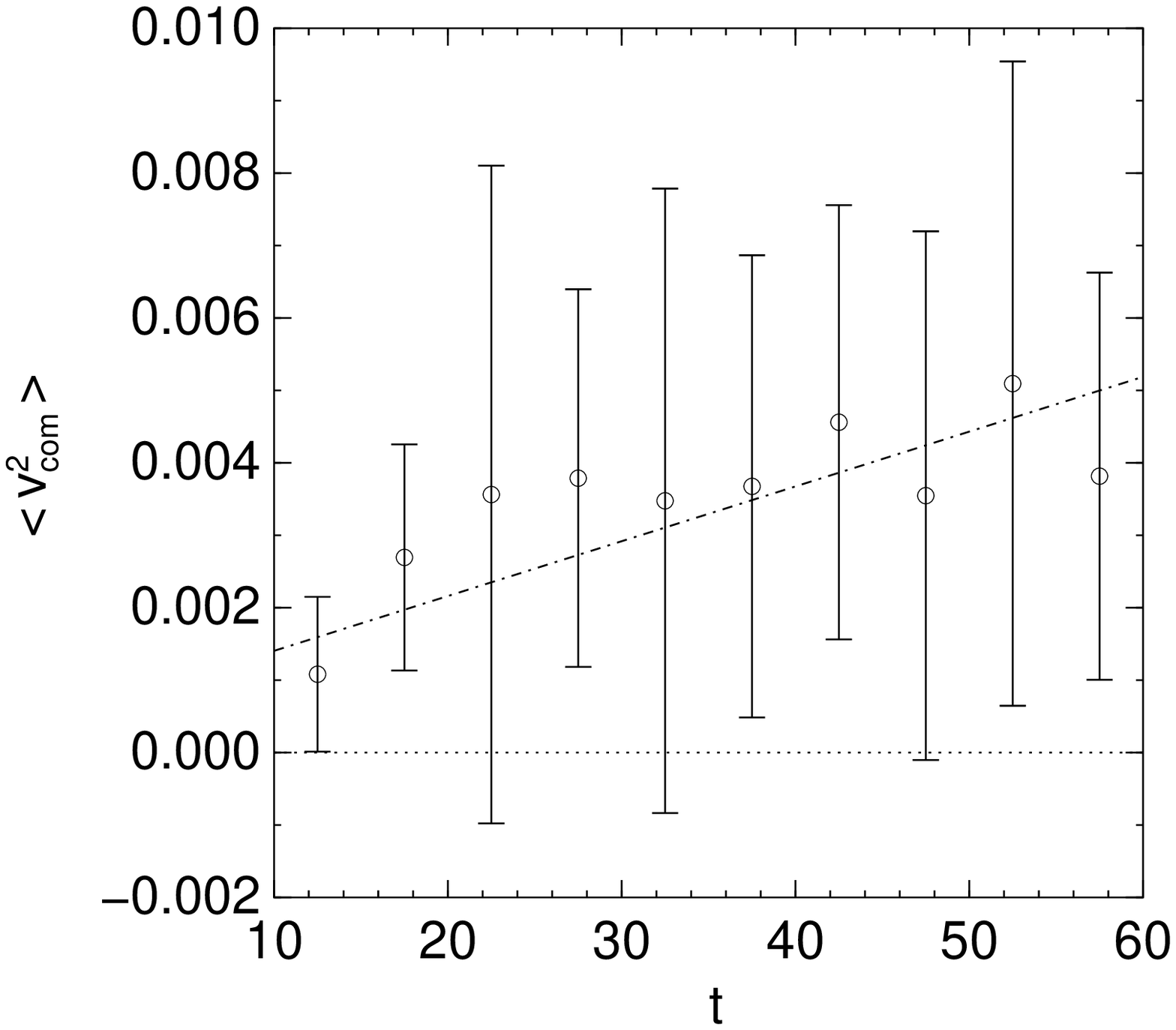}
\caption{The mean of $v^2_\mathrm{com}$ as a function of the
integrated time in $N$-body units. The left plot shows results for the
differen particle groups: 32768 data are plotted with open triangles,
65536 data with open squares, and 131072 data with circles. In the
right plot the data for $v^2_\mathrm{com}$ have been averaged over all
particle groups, the error-bars represent the standard deviation in
the data.}
\label{fig:barvsqcomtime}
\end{figure}

\begin{figure}
\plotone{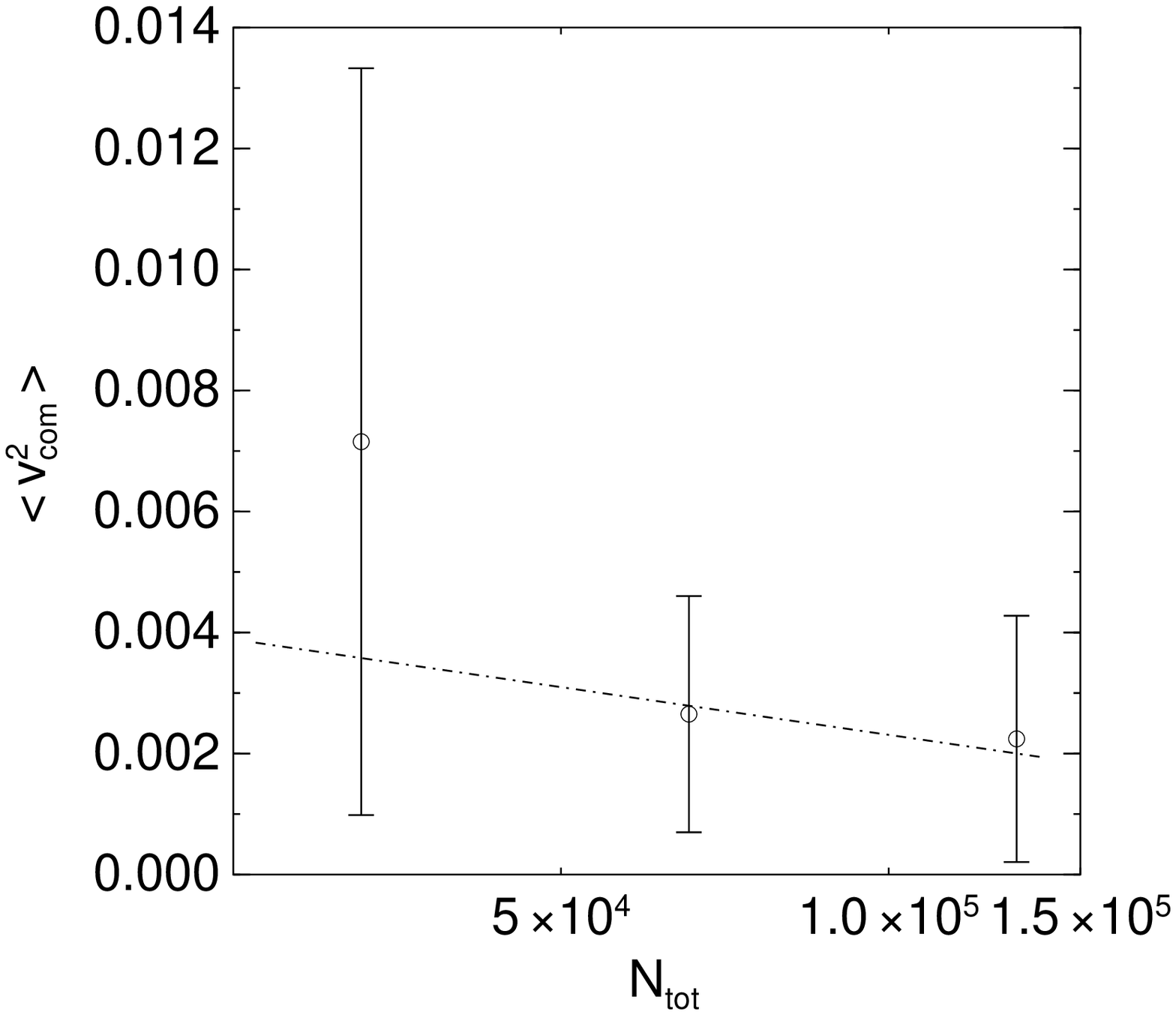}
\caption{The mean of $v^2_\mathrm{com}$ as a function of the total particle
numbers in the simulations. The values for $v^2_\mathrm{com}$ have
been averaged over the total simulation time, the error-bars represent
the standard deviation in the data. $v_\mathrm{com}$ is the relative
motion of the center of mass of the massive binary relative to the
center of mass of the stellar system.}
\label{fig:barvsqcomntot}
\end{figure}

Figures \ref{fig:barvsqcomntot} and \ref{fig:barvsqcomtime} show
the evolution of $\langle v^2_\mathrm{com} \rangle$ as a function of
time and total particle number $N_\mathrm{tot}$. The quantity
$v_\mathrm{com}$ is the velocity of the center of mass of the black
holes relative to the velocity of the center of mass of the stellar
system. For the time dependence of $\langle v^2_\mathrm{com} \rangle$,
we find a slope of $(7.6 \pm 4.9) \times 10^{-5}$ for the fitting line
in Figure \ref{fig:barvsqcomtime}. For differing total particle
numbers this slope is $(3.5 \pm 7.0) \times 10^{-5}$ for 32768
particles, $(2.6 \pm 2.4) \times 10^{-5}$ for 65536 particles, and
$(4.2 \pm 4.0) \times 10^{-5}$ for 131072 particles. The slope for the
dependence of $\langle v^2_\mathrm{com} \rangle$ on particle number in
Figure \ref{fig:barvsqcomntot} is $(-1.7 \pm 3.9) \times 10^{-8}$.

\subsubsection{Connection between the wandering and the orbital decay}

\begin{figure}
%\plottwo{f15.eps}{f16.eps}
\plottwo{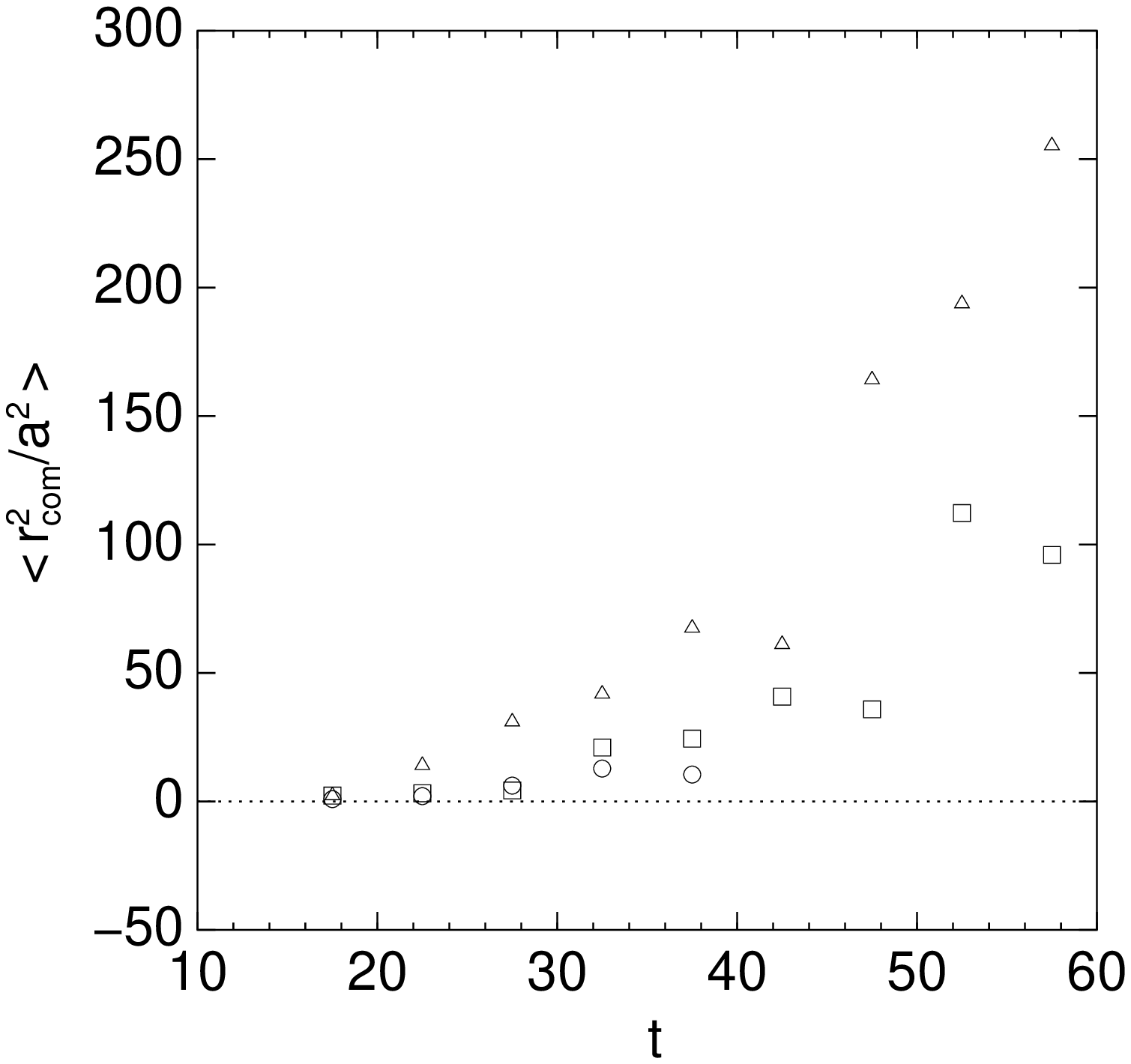}{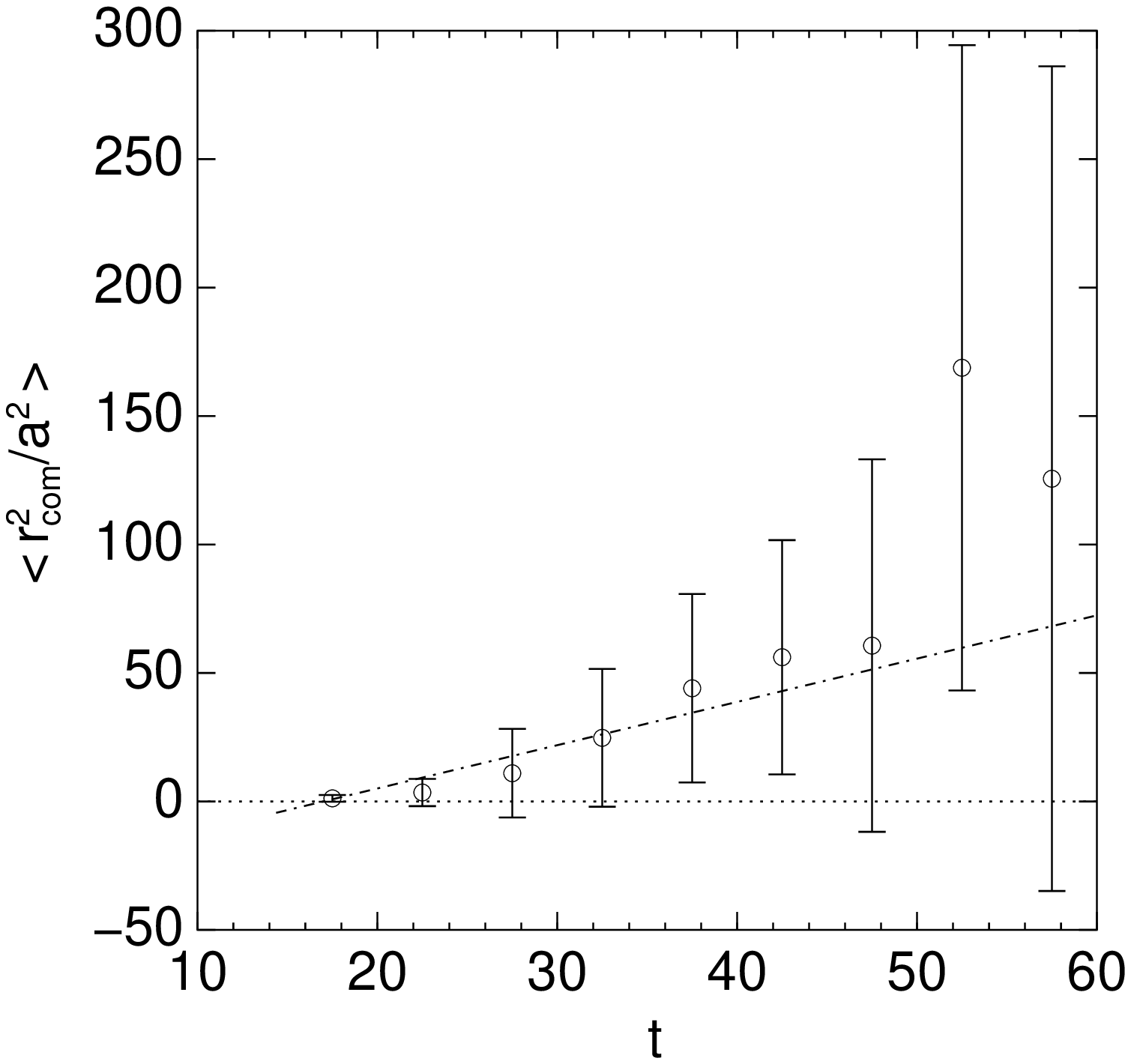}
\caption{Wandering of the binary in relation to the squared semi-major
  axis $a^2$ of the bound black hole binary as a function of time. The
  left plot shows the results for each particle group: 32768 data are
  plotted with open triangles, 65536 data with open squares, and
  131072 data with circles. The right plot shows the results for all
  simulations put together.}
\label{fig:sinkwandering}
\end{figure}

Figure \ref{fig:sinkwandering} shows the ratio of the wandering and
the semi-major axis of the binary orbit $a^2$ as a function of time. The
evolution of this ratio has a strong dependence on the particle
number, as the wandering is dependent on the simulation size. However,
all simulations show the same trend in that wandering becomes more
important with time for the binary. As the right plot in Figure
\ref{fig:sinkwandering} shows, a fitting line with a slope of $1.7 \pm
0.6$ can fit the data. However, the data suggests a nonlinear
behavior which should be roughly quadratic, since $\langle 1/a
\rangle$ increases linearly and $\langle r^2_\mathrm{com} \rangle$ is
roughly constant. 

\subsubsection{The effect of dynamical friction}

\begin{figure}
%\includegraphics[width=\linewidth]{Figures/ll-tim-all.eps}
%\plottwo{f18.eps}{f9.eps}
\plottwo{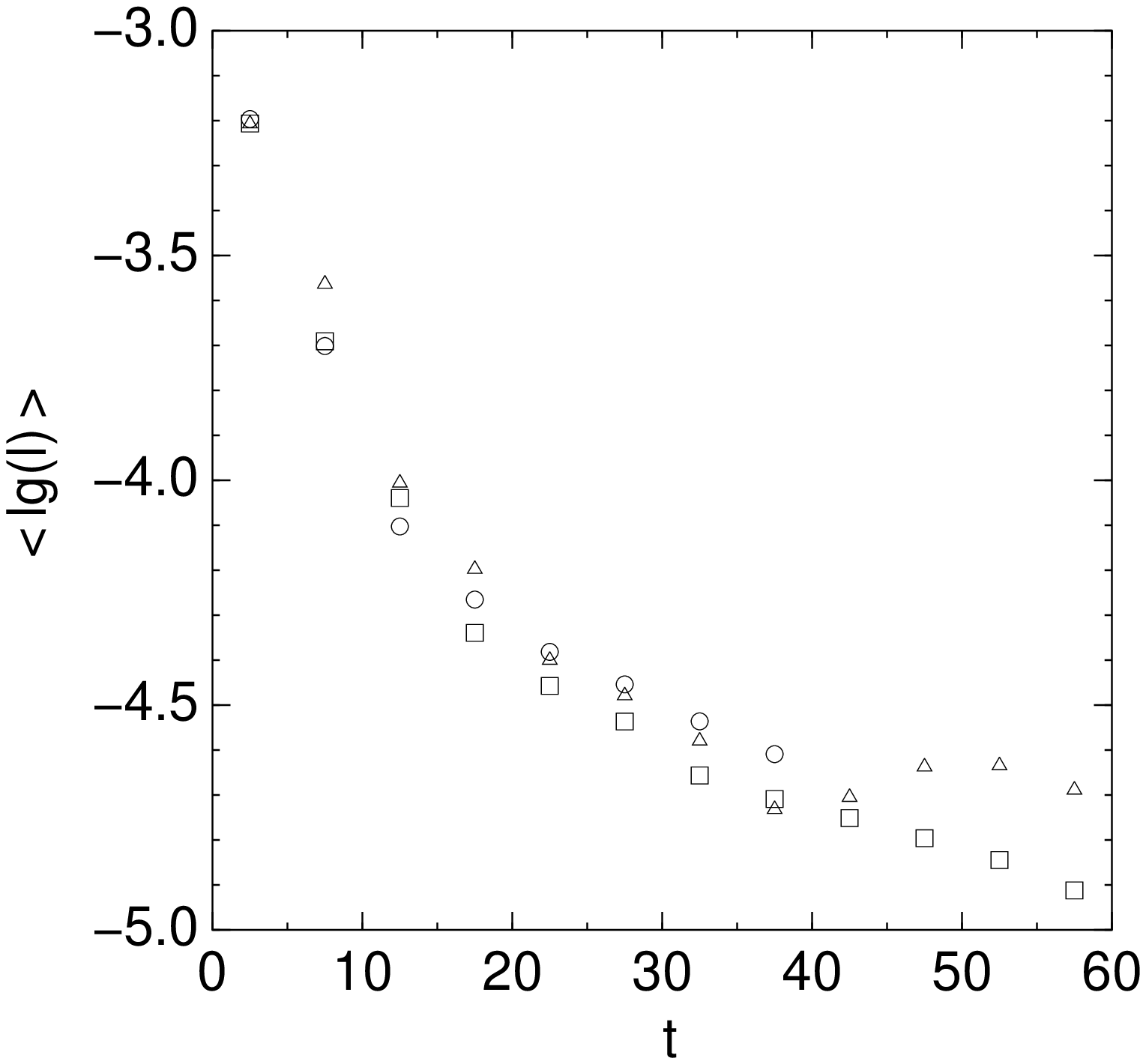}{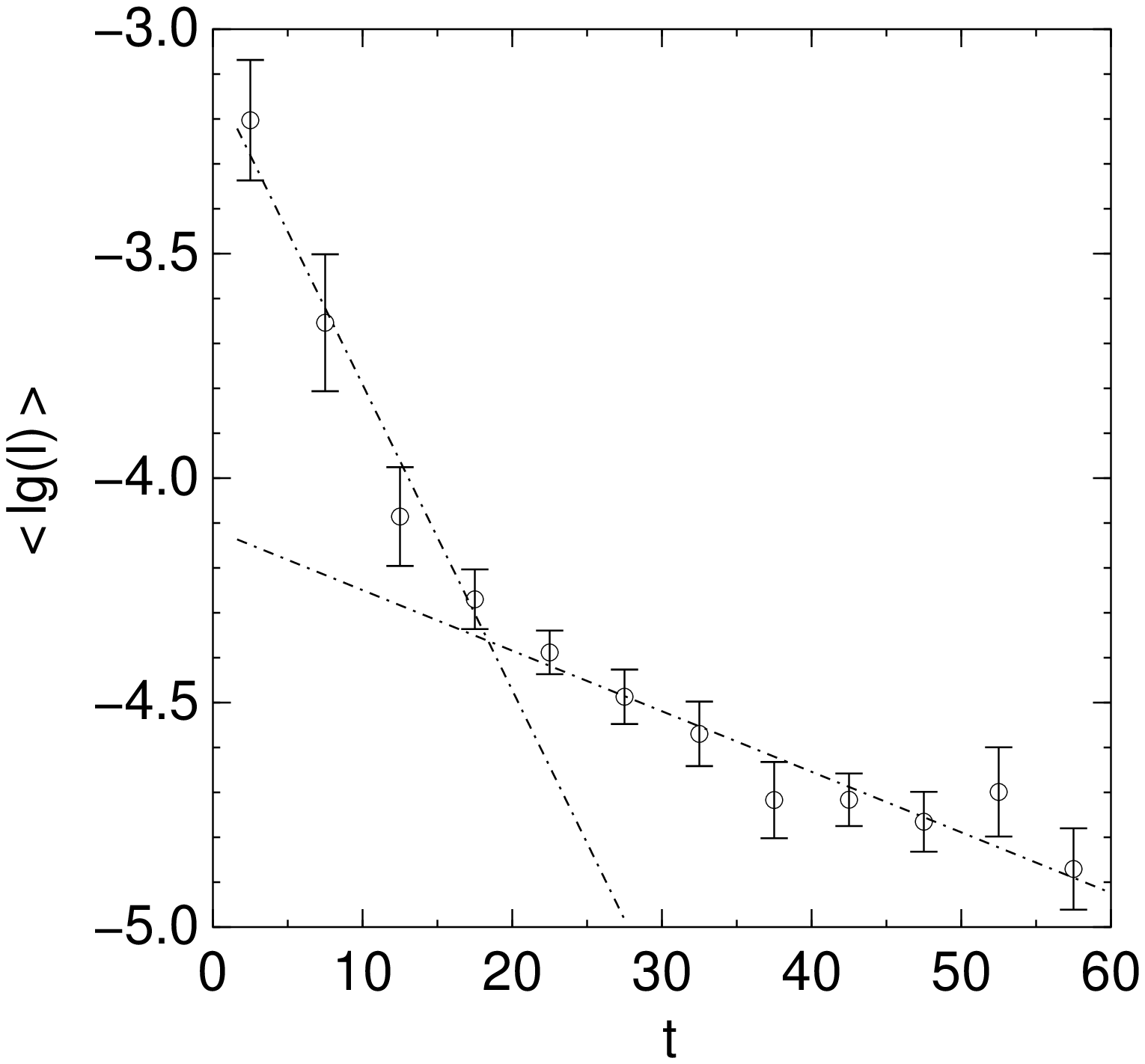}
\caption{The evolution of the orbital angular momentum as a function
of time for the collected data of the runs. The left plot shows the
averages computed for each particle group. 32768 data are plotted with
open triangles, 65536 data with open squares, and 131072 data with
circles.  The right plot shows the averages for all particle groups
together. The error bars represent the standard deviation in the
data. In order to distinguish between the two modes of evolution,
linear regression is applied to the bins between 0 and 20 and 20 and
60 time units separately.}
\label{fig:llalltime}
\end{figure}

To study the influence of dynamical friction on the decay of the
binary orbit we analyze the behavior of its orbital angular momentum
as a function of time. As Figure \ref{fig:llalltime} shows, the decay
shows a two mode evolution. Between 0 and 20 time units, linear
regression for $\langle \lg(l) \rangle$ gives a slope of $(-6.8 \pm
0.9) \times 10^{-2}$. The line with the more shallow slope $(-1.3 \pm
0.2) \times 10^{-2}$ represents the behavior between 20 and 60 time
units.

\subsection{Reaction of the stellar system}

\begin{figure}
%\includegraphics[width=\linewidth]{Figures/rho0-tim-all.eps}
%\plotone{f10.eps}
\plotone{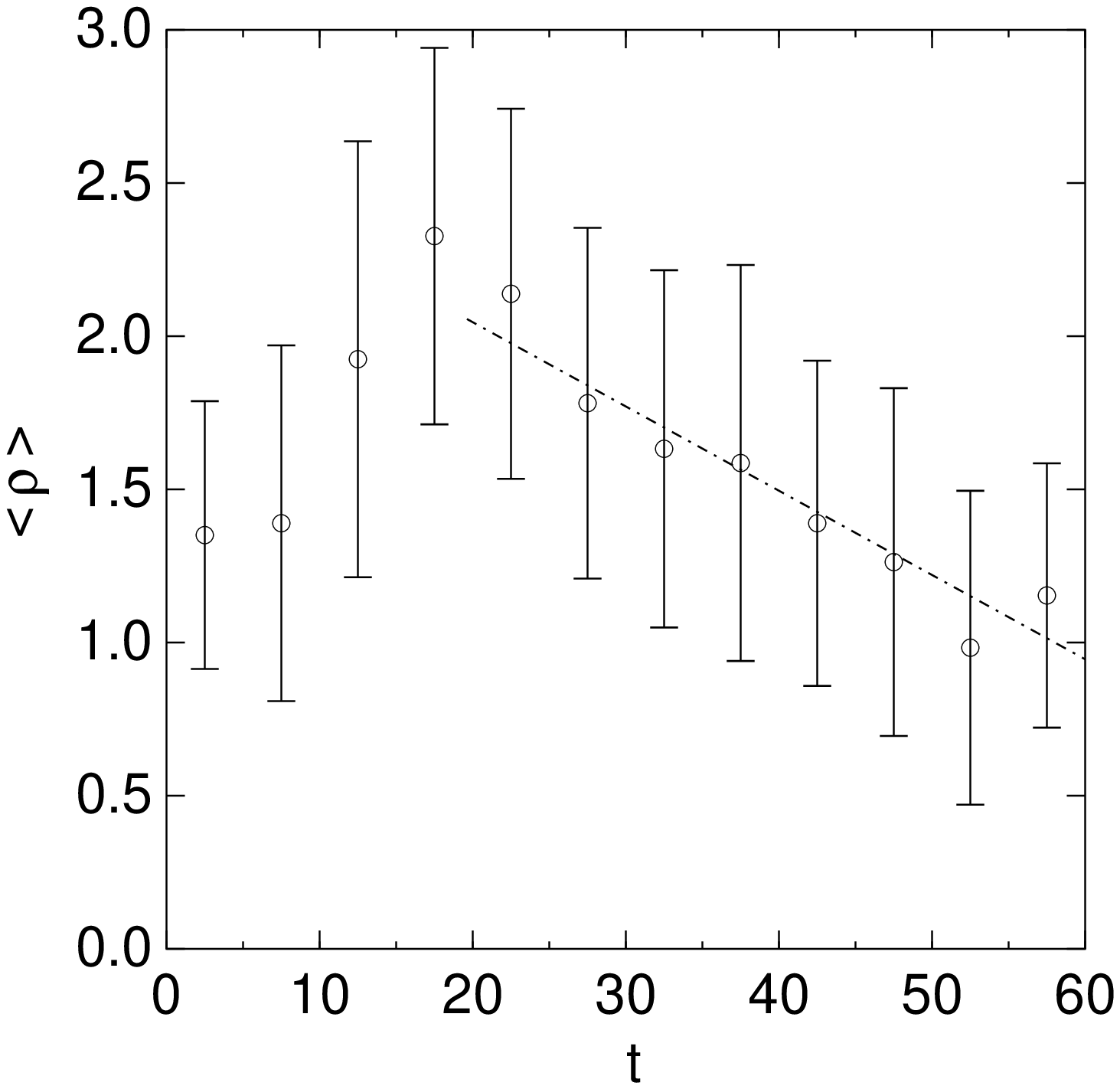}
\caption{Evolution of the stellar density in a central sphere of the
cluster with $r_\mathrm{csp} = 0.032$. $\rho$ is the average over all
simulations, the error-bars represent the standard deviation in the
data. The dot-dashed line shows our linear fit for the evolution
between 20 and 60 time units.}
\label{fig:rhoinner}
\end{figure}

\begin{figure}
%\includegraphics[width=\linewidth]{Figures/sigma0-tim-all.eps}
%\plotone{f11.eps}
\plotone{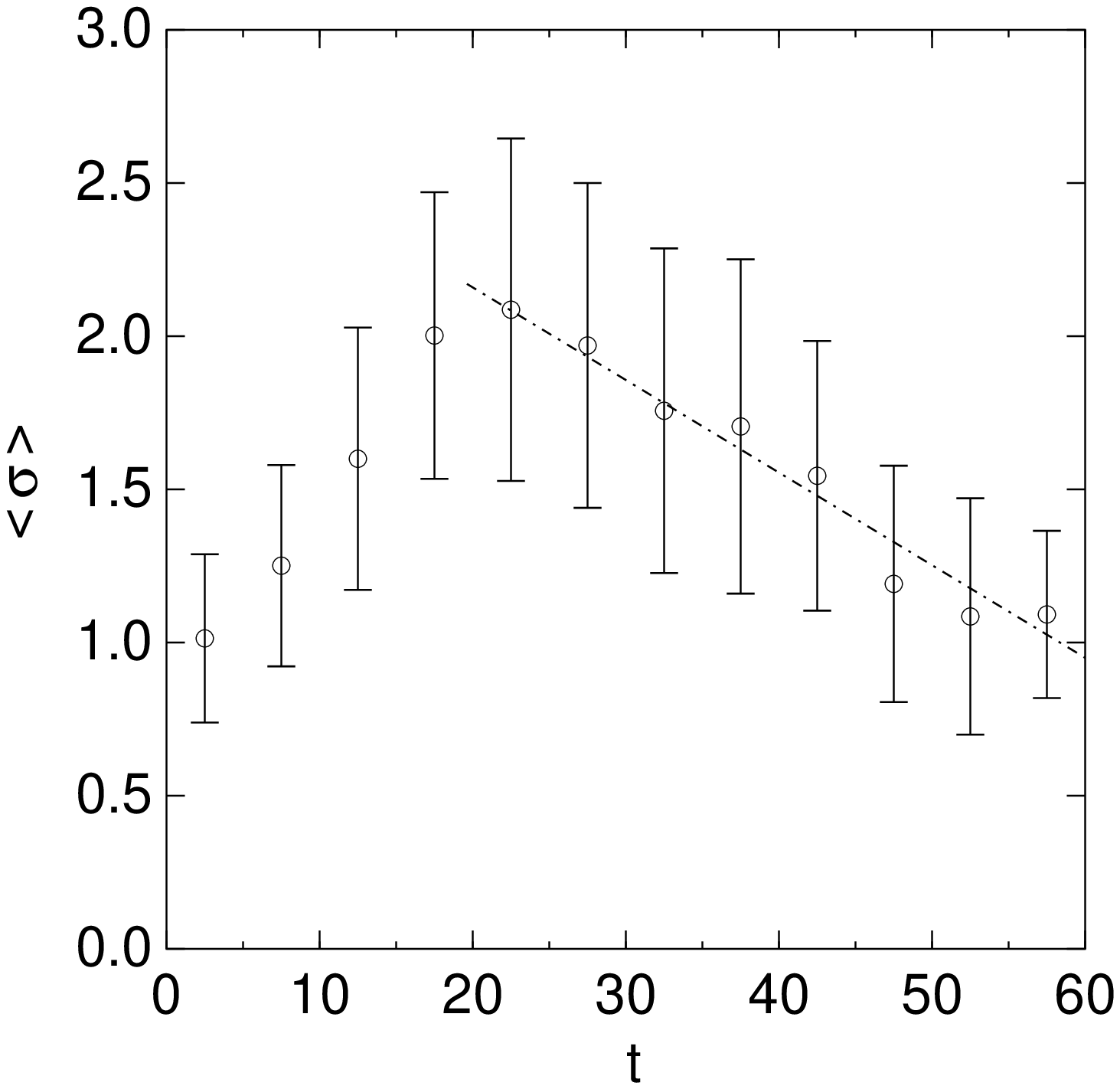}
\caption{Evolution of the stellar velocity dispersion in a central
sphere of the cluster with $r_\mathrm{csp} = 0.032$. The quantity
$\sigma$ is the average from all simulations, the error-bars show the
standard deviation in the data. The dot-dashed line shows our linear
fit for the evolution between 20 and 60 time units.}
\label{fig:sigmainner}
\end{figure}

The stellar system reacts to the motion of the black hole in a generic
fashion. We find that our statistical basis is too small for finding a
clear dependency of the results on the total number of particles in
the simulations. Hence we present only the averages from all of our
runs.  Figures \ref{fig:rhoinner} and \ref{fig:sigmainner} show the
evolution of the density and the velocity dispersion respectively for
particles within a radius of $r_\mathrm{csp} = 0.032$ averaged over
all runs as a function of time. While the black hole binary becomes
bound at $\approx 10$ time units, the density has a maximum at
$\approx 18$ time units, and the velocity dispersion is highest at
$\approx 23$ time units. A linear fit ($y = a + bx$) has been applied
to the evolution of $\rho$ and $\sigma$ between 20 and 60 time units
as plotted in Figures \ref{fig:rhoinner} and \ref{fig:sigmainner}. For
$\langle \rho \rangle$, we find $a = 2.6 \pm 0.7$ and $b = -0.028 \pm
0.016$, for $\langle \sigma \rangle$ we find $a = 2.8 \pm 0.6$ and $b
= -0.030 \pm 0.013$.

% WHAT DO THESE FITTED NUMBERS MEAN PHYSICALLY? (ERIC)

\section{Discussion}

\subsection{Hardening rate}
Following \citet{Hills:92}, and \citet{Quinlan:96} the hardening rate
$H$ of a massive binary floating in a sea of light stars is given by,
\begin{equation}
\frac{d}{dt} \, \frac{1}{a} = H \, \frac{G \rho}{\sigma}.
\label{eq:hardening}
\end{equation}
With $G = 1$ and the assumption that the averages for $\rho$ and
$\sigma$ evolve in the same way between 20 and 60 time units, which
would render the ratio between $\rho$ and $\sigma$ constant, we find
$H = 8.7 \pm 0.4$. This is significantly smaller than the values given
by \citet{Hills:92} ($H = 13.5$) and \citet{Quinlan:96} ($H \approx
18$).

Our smaller hardening rate compared to the results of
\citet{Quinlan:96} is caused by the lower central density and the core
type radial density profile of our Plummer model. \citet{Quinlan:96}
uses Jaffe models for his simulations which allow rapid transfer of
orbital energy into the dense cusp through tidal interactions. This is
also represented in the destruction of the cusp \citet{Quinlan:96}
observes, while our simulations show a much weaker change for the
central density. 

\citet{Hills:92} models the shrinking of the binary orbit through
three body encounters. His greater value of $H$ is consistent with our
simulations. We observe steplike changes of the binding energy at
later times of the simulations, which is less pronounced with
increasing particle numbers. Because the granularity of the potential
is higher in low $N_\mathrm{tot}$ runs, three body interactions with
the black hole binary become more likely. As shown in Figure
\ref{fig:esnb6bin-epshh} such three body encounters can enhance the
decay of the orbit. Thus, our small $H$ indicates that in our
simulations shrinking of the black hole orbits is mainly caused by
dynamical friction and not so much by tidal destruction of cusps or
three body encounters.

\subsection{Brownian motion}
If the black holes reach equipartition of kinetic energy with the
stars, their expected mean square velocity follows from,
\begin{equation}
\langle v_\mathrm{equ}^2 \rangle = \frac{m_*}{m_\mathrm{com}} \, 
	\langle v_*^2 \rangle.
\label{eq:equipartition}
\end{equation}
$\langle v_\mathrm{equ}^2 \rangle$ is the mean square velocity we
expect for a particle with mass $m_\mathrm{com}$, which is the
combined mass of the two black holes. $m_*$ is the mass of the stars
and $\langle v_*^2 \rangle$ their mean square velocity.
Since our setup involves a Plummer model, we are expecting the binary
to move in a harmonic potential in later stages of a simulation and
for $\langle r_\mathrm{com}^2 \rangle \propto \langle v_\mathrm{com}^2
\rangle$.

While the individual black holes do not reach equipartition, equation
(\ref{eq:equipartition}) can describe the Brownian motion of the
system. The sum of the black hole masses is represented by
$m_\mathrm{com}$ and the center of mass motion of the binary by
$v_\mathrm{com}$. The mass ratios between the individual stars and the
black hole binary are $1.49 \times 10^{-3}$, $7.47 \times 10^{-4}$,
and $3.74 \times 10^{-4}$ in the runs with 32768, 65536, and 131072
particles, respectively. However, equation (\ref{eq:equipartition})
does not describe the behavior of the center-of-mass motion
correctly. While the velocity dispersion drops after 20 time units,
the mean square velocity of the black holes $\langle v^2_\bullet
\rangle$ increases. For this reason, we compare the measured average
$\langle v_\mathrm{com}^2 \rangle$ for the datasets with the mean
expectation from the right side of equation
(\ref{eq:equipartition}). For $\langle v_*^2 \rangle$, we take the
average over all simulations, which is $1.5$, neglecting the
variability over time. Using this, we can estimate $\langle
v_\mathrm{equ}^2 \rangle$ and compare it with the measured $\langle
v_\mathrm{com}^2 \rangle$,

\begin{tabular}{cccc}
$N_\mathrm{tot}$ &   $\langle v_\mathrm{com}^2 \rangle$ & 
$\langle v_\mathrm{equ}^2 \rangle$ & 
$\langle v_\mathrm{com}^2 \rangle/\langle
v_\mathrm{equ}^2 \rangle$ \\ \hline
32768 & 0.0072 & 0.0022 & 3.2 \\
65536 & 0.0026 & 0.0011 & 2.3 \\
131072 & 0.0022 & 0.0006 & 3.9
\end{tabular}

This means we find a center of mass motion for the binary which
exceeds the expected value from Brownian motion. However, the motion
is enhanced by larger factors than proposed by \citet{Merritt:2000-b}
or by \citet{Chatterjee:2002}. A more detailed discussion of this
result remains for future work.

\subsection{Dynamical friction}
Following \citet{Begelman:80}, dynamical
friction becomes inefficient as the driving forces behind the binary black
hole orbit decay after it becomes hard. In
order to put constraints on this, we estimate the influence of
dynamical friction on the decay of the binary in the simulation. From
\citet{Binney:87}, we take the following
expression, which is derived in \citet{Chandrasekhar:43-a}.
\begin{equation}
\frac{d\mathbf{v}_\bullet}{dt} = - 16 \pi^2 G^2 \, \log(\Lambda) \,
	M_* (M_\bullet + M_*) \, \frac{\int\limits_0^{v_\bullet} \,
	v_*^2 f(v_*,\mathbf{r}) dv_*}{v_\bullet^3} \, \mathbf{v}_\bullet.
\label{eq:dynbinney}
\end{equation}
Equation (\ref{eq:dynbinney}) describes the deceleration of a particle
with mass $M_\bullet$ under the influence of weak encounters with
surrounding particles having a uniform mass $M_*$. In the special case
of a Plummer model, the integral in equation (\ref{eq:dynbinney}) can
be expressed in terms of the escape velocity $v_\mathrm{esc}$
\citep{Aarseth:74},
\begin{gather}
\int\limits_0^{v_\bullet} \, v_*^2 f(v_*,\mathbf{r}) dv_* = 
	\frac{n(\mathbf{r})}{C} \,
	\int\limits_0^{q_\bullet} q^2 \, (1 - q^2)^{\frac{7}{2}} dq,
	\label{eq:myplum} \\
\intertext{where,}
C = \int\limits_0^1 q^2 \, (1 - q^2)^{\frac{7}{2}} dq.
\end{gather}
The quantity $n(\mathbf{r})$ defines the number density of the stars
at the position $\mathbf{r}$ and $q = v/v_\mathrm{esc}$. Taking the
limit of a continuous system with $M_* \ll M_\bullet$, the term
$n(\mathbf{r}) M_* (M_\bullet + M_*)$ becomes $M_\bullet
\rho(\mathbf{r})$. The integral over $q$ in equation (\ref{eq:myplum})
can then be solved in a closed form.

If the motion of the black holes is determined by their self
interaction plus a frictional force term, this friction can be linked
to the decay of the angular momentum $l$ as follows:
\begin{gather}
\mathbf{a} = \mathbf{a}_r + a_\mathrm{df} \frac{\mathbf{v}}{v},\\
\dot{l} = \frac{m a_\mathrm{df}}{v} \, ( \mathbf{r} \times \mathbf{v}
) \; = \; \frac{l a_\mathrm{df}}{v},
\label{eq:dotlzul}
\end{gather}
where $\mathbf{a}_r$ is the radial acceleration of the two body motion
of the black holes, $a_\mathrm{df}$ is the dynamical friction acting
on each black hole, and $\mathbf{v}$ is the two-body velocity of the
black holes. With equations (\ref{eq:dynbinney}) and (\ref{eq:myplum})
we can evaluate the impact of dynamical friction on the orbital
angular momentum according to
\begin{equation}
\frac{\dot{l}}{l} = - \frac{16 \pi^2 G^2}{C v_\bullet^3} \,
	\log(\Lambda) M_\bullet \rho(\mathbf{r})
	\int\limits_0^{q_\bullet} q^2 \, (1 - q^2)^{\frac{7}{2}} dq.
\label{eq:dynmarc}
\end{equation}
The gravitational constant $G$ is unity in our model units, and the
black holes have mass $M_\bullet = 0.01$ each. We use the mean orbital
velocities for $v_\bullet$. In order to evaluate $\rho(r)$, we use the
mean separation of the black holes for $r$, which introduces only a
very small error in a Plummer model. Assuming a linear behavior for
$\lg(l) = a + bt$, we find $\dot{l}/l = b \ln(10)$. Using this to
estimate the angular momentum from the slopes $b$ of the linear fits
in Figure \ref{fig:llalltime}, we find $\log(\Lambda) \approx
0.15$. This result shows that the usual assumption of large $\Lambda$
does not hold.

Both the possibility of a linear fit for the evolution of $l$ and the
small hardening rate $H$ indicate that mainly dynamical friction
causes the shrinking of black hole orbits in our simulations.

\section{Physical units}

As stated before, the collisional simulations which include black hole
particles do not reach the observed mass contrast in galactic
nuclei. In order to transform simulation units to physical units, a
system size in parsec or a stellar mass in units of solar mass has
to be chosen. Setting the gravitational constant $G$, all
remaining units can be rescaled \citep{Heggie:86}.

In the following, a run with 65536 particles is scaled to a physical
stellar system.  Since this work focuses on the dynamics of galactic
nuclei, the physical mass of the supermassive objects motivates the
following choices,
\begin{gather}
M_\bullet \equiv 1.00015 \times 10^7 \, \mathrm{M}_\odot, \\
M_* \equiv 1.49536 \times 10^4 \, \mathrm{M}_\odot.
\end{gather}
The masses are chosen so that the total mass of the system is
$M_\mathrm{tot} = 10^{9} \, \mathrm{M}_\odot$ and the mean mass of a
particle is $\bar{M} = 1.52588 \times 10^4 \, \mathrm{M}_\odot$.  This
choice means that every stellar particle with mass $M_*$ represents a
compact star cluster with the order of $10^4$ particles. The chosen
mass for the black hole particle has approximately the same mass as
the central black hole of M31 \citep{Magorrian:98}.

The conversion between physical units and
$N$-body units follows $x_\mathrm{phys} =
X_\mathrm{conv} x_\mathrm{sim}$ for simulated quantities.
By choosing the central velocity dispersion to be $110 \;
\mathrm{km}/\mathrm{s}$, we find $T_\mathrm{conv}$ and
$R_\mathrm{conv}$,
\begin{gather}
R_\mathrm{conv} = 355.39 \, \mathrm{pc}, \\
M_\mathrm{conv} = 10^9 \, \mathrm{M}_\odot, \\
T_\mathrm{conv} = 3.1590 \times 10^6 \, \mathrm{y}, \\
V_\mathrm{conv} = 110 \, \mathrm{km}/\mathrm{s}.
\end{gather}
In a Plummer model, the half mass radius $r_h$ is related to the
Plummer radius $R$ by $r_h = 1.30 R$ \citep{Spitzer:87}. Since $R = 3
\pi/16$, the half mass radius in model units is $r_h = 0.766$.
Therefore, the initial model for the simulated decay of a black hole
in the galactic center is a Plummer sphere with a half mass radius of
272.23 pc. The initial distance between the black holes is 355.39
pc. They become bound after approximately 40 million years. The total
simulated time is approximately 190 million years. At the end of the
simulation the black hole distances vary from 1 pc at
apocenter to 0.2 pc at pericenter. The semi-major axis of the
first bound orbit is 21 pc. 

This scaling allows us to compare our results with
\citet{Begelman:80}. We find that our smallest average orbits at the
end of the simulation are not yet small enough that gravitational
radiation, according to their estimates, would dominate the evolution
time scale. 

However, at the end of our simulations evolution is still dominated by
dynamical friction and not by long evolution time scales for hard
binaries as proposed by \citet{Begelman:80}. Their estimate for the
gravitational radiation shrinking time scale assumes circular orbits
for the binary. With eccentricities of roughly 0.85 for the black hole
binaries in our runs, we expect gravitational radiation to be efficient
and coalescence in roughly $10^8$ years after our simulations stopped.

\section{Conclusions}

We have created a new $N$-body hybrid code by merging a high accuracy
direct Hermite integrator of the standard type
\citep{Aarseth:99,Spurzem:99} with a collisionless $N$-body method
which approximates the potential of a given particle distribution by a
series expansion \citep{Hernquist:92,Zhao:96}. The SCF method has been
completely rewritten to include a computation of the time derivative
of the gravitational force and a fourth order Hermite integrator.  We
have used this code to model a galactic nucleus containing two massive
black holes with up to 128k single particles. The evolution of the
binary black hole is followed from an initial phase, to a phase driven
by standard dynamical friction where the binary is bound, and then
further hardened by three-body encounters with single stars. In that
hardening phase, we take full advantage of the regularized three-body
integration developed by \citet{Mikkola:96} and
\citet{Mikkola:98}. The method proves to work well, and reproduces
standard expectations, such as the Chandrasekhar dynamical friction in
the initial phase.  In the final hardening phase due to three-body
encounters, we find that the eccentricity of the black hole binary
maintains a fairly large value (around 0.85). This is very interesting
because it decreases the time scale for gravitational radiation merger
of binary black holes dramatically, thus increasing our chances of
detecting gravitational radiation from such events with LISA. Due to
computational limitations, however, our particle numbers are still not
large enough to fully describe the real physical situation. Any
further scaling is problematic, and so further work with improved
hardware and software must be done.

We study in detail the motion of a black hole binary in the center of
a galaxy. We find that the wandering motion does not decay with
increasing particle number as expected. The mechanism exciting these
anomalous motions is unclear. If they exist in simulations with
realistic particle numbers, they will solve the problem of feeding the
black holes with fresh stellar dynamical material raised by
\citet{Gould:2000}.

\section{Acknowledgements}

The authors would like to thank S.~Aarseth, D.~Heggie, W.~Sweatman,
C.~Theis, C.~Boily, D.~Merritt, M.~Milosavljevi\'c, F.~Cruz,
H.~Baumgardt, G.~Hensler, L.~Hernquist, H.~S.~Zhao, P.~Ghavamian and
E.~Barnes for fruitful help and discussion.  This project is funded by
\emph{Deutsche Forschungsgemeinschaft} (DFG) project Sp 345/9-1,2 and
Sonderforschungsbereich (SFB) 439 funded at the University of
Heidelberg, NSF grant AST 00-71099, NASA grants NAG 5-7019, NAG
5-6037, and NAG 5-9046.  Technical help and computer resources are
provided by \emph{NIC} in J\"ulich, \emph{HLRS} in Stuttgart,
\emph{TRACS} and \emph{EPCC} in Edinburgh, \emph{ZIB} in Berlin,
\emph{SSC} in Karlsruhe, University of Heidelberg, Rutgers University,
and University of Kiel, and by the Pittsburgh Supercomputer Center and
the San Diego Supercomputer Center.  The authors thank the Aspen
Center for Physics, the Institute of Astronomy and the Lorenz center
at Leiden University for hospitality.  The sources for EuroStar are
available from the authors or via
\verb+http://www.physics.rutgers.edu/~marchems/+

\appendix

\section{Recurrence relations for Ultraspherical polynomials}

\label{kap:Rekursion}
Throughout the computation for the forces and force derivatives in our
SCF-scheme several special functions have to be tabulated. Recurrence
relations provide a very efficient means of calculating these
functions. The following recursion relations have been applied to
compute ultraspherical polynomials and their derivatives. As starting
values for $n \in {0, 1}$ the recurrence formulae for the Gegenbauer
or ultraspherical polynomials obey the relation
\begin{equation}
C_n^{(\alpha)}(\xi) = 
\begin{cases}
1 & \text{if}\quad  n = 0, \\
2 \alpha \xi & \text{if} \quad n = 1.
\end{cases}
\end{equation}
The expressions for higher values are given by:
\begin{equation}
C_{n + 1}^{(\alpha)}(\xi) = \frac{2 \, (n + \alpha) \, \xi \,
	C_n^{(\alpha)}(\xi) - (n + 2 \alpha - 1) \, C_{n -
	1}^{(\alpha)}(\xi)}{(n + 1)}
	\label{eq:ultrasp} 
\end{equation}
From that the first derivative can be computed as:
(\citet{Abramowitz:72}, equation (22.7.22) and table 22.7.) 
\begin{equation}
C_{n - 1}^{(\alpha + 1)}(\xi) = \frac{(n + 2 \alpha) \, \xi \,
	C_n^{(\alpha)}(\xi) - (n + 1) \, C_{n + 1}^{(\alpha)}(\xi)}{2
	\alpha \, (1 - \xi^2)}
	\label{eq:ultrasp1}
\end{equation}
For practical reasons and higher accuracy the second derivative
polynomial is computed using equation (\ref{eq:ultrasp}):
\begin{equation}
C_{n + 1}^{(\alpha + 2)}(\xi) = \frac{2 \, (n + \alpha + 2) \, \xi \,
	C_n^{(\alpha + 2)}(\xi) - (n + 2 (\alpha + 2) - 1) \, C_{n -
	1}^{(\alpha + 2)}(\xi)}{(n + 1)}
	\label{eq:ultrasp2} 
\end{equation}

\label{kap:GradPhiDot}

Because the particle track is approximated by using the Hermite
scheme, one has to find forces and the first force derivative
simultaneously. An approximation using two timesteps for the first
force derivative introduces errors to the second and third derivative
of the forces. All particles move within a time dependent potential;
therefore, the first derivative has a term describing the change of
the potential and a term describing the change of force depending on
the particle's orbit.
\begin{equation}
\frac{d}{dt} \, \mathbf{a}(t,\mathbf{r}) = \frac{\partial
	\mathbf{a}(t,\mathbf{r})}{\partial t} + \frac{\partial 
	\mathbf{r}}{\partial t} \, \frac{\partial
	\mathbf{a}(t,\mathbf{r})}{\partial \mathbf{r}}.
	\label{eq:SCFadot}
\end{equation}
With the help of the orbit integration for the single particle in a given
static potential case, equation (\ref{eq:SCFadot}) evaluates
to:
\begin{equation}
\begin{split}
\frac{d}{dt} \, \mathbf{a}(t,\mathbf{r}) &= ( \frac{\partial a_r}{\partial
	t} + \frac{\partial a_r}{\partial r} \dot{r} +
	\frac{\partial a_r}{\partial \vartheta} \dot{\vartheta} +
	\frac{\partial a_r}{\partial \varphi} \dot{\varphi} -
	a_\vartheta \dot{\vartheta} - a_\varphi \dot{\varphi}
	\sin{\vartheta} ) \, \mathbf{e}_r \\
	&\quad + ( \frac{\partial a_\vartheta}{\partial t} +
	\frac{\partial a_\vartheta}{\partial r} \dot{r} +
	\frac{\partial a_\vartheta}{\partial \vartheta}
	\dot{\vartheta} + \frac{\partial a_\vartheta}{\partial
	\varphi} \dot{\varphi} + a_r \dot{\vartheta} - a_\varphi
	\dot{\varphi} \cos{\vartheta} ) \, \mathbf{e}_\vartheta \\
	&\quad + ( \frac{\partial a_\varphi}{\partial t} + 
	\frac{\partial a_\varphi}{\partial r} \dot{r} +
	\frac{\partial a_\varphi}{\partial \vartheta}
	\dot{\vartheta} + \frac{\partial a_\varphi}{\partial
	\varphi} \dot{\varphi} + a_r \dot{\varphi} \sin{\vartheta} +
	a_\vartheta \dot{\varphi} \cos{\vartheta} ) \,
	\mathbf{e}_\varphi 
	\label{eq:SCFadotorb}
\end{split}
\end{equation}

The evalutation of the first term on the right hand side of equation
(\ref{eq:SCFadot}) is given in section \ref{kap:GradPhiDotTim}. The
derivatives with respect to the spatial coordinates in equation
(\ref{eq:SCFadotorb}) can be found in section \ref{kap:GradPhiDotOrb}.

\section{Time-dependency of the potential}
\label{kap:GradPhiDotTim}

Because all positions and velocities of the dataset are
time-dependent, the partial derivatives with respect to $t$ apply only
to the coefficients $A_{nlm}$. These are implemented as the variables
$C_{lm}(r)$, $D_{lm}(r)$, $E_{lm}(r)$, and $F_{lm}(r)$, from which the
partial derivative can be formed:
\begin{gather}
\frac{\partial C_{lm}(r)}{\partial t} = N_{lm} \, \sum\limits_{n =
	0}^{\infty} \, 
	\tilde{A}_{nl} \tilde{\Phi}_{nl}(r) \, \sum\limits_k \, m_k
	\frac{\partial}{\partial t} \left( \tilde{\Phi}_{nl}(r_k) \,
	P_{lm}(\cos(\vartheta_k)) \, \cos(m \varphi_k) \right), \\
\frac{\partial D_{lm}(r)}{\partial t} = N_{lm} \, \sum\limits_{n =
	0}^{\infty} \, 
	\tilde{A}_{nl} \tilde{\Phi}_{nl}(r) \, \sum\limits_k \, m_k
	\frac{\partial}{\partial t} \left( \tilde{\Phi}_{nl}(r_k) \,
	P_{lm}(\cos(\vartheta_k)) \, \sin(m \varphi_k) \right), \\
\frac{\partial E_{lm}(r)}{\partial t} = N_{lm} \, \sum\limits_{n =
	0}^{\infty} \, 
	\tilde{A}_{nl} \, \frac{d \tilde{\Phi}_{nl}(r)}{dr} \,
	\sum\limits_k \, m_k 
	\frac{\partial}{\partial t} \left( \tilde{\Phi}_{nl}(r_k) \,
	P_{lm}(\cos(\vartheta_k)) \, \cos(m \varphi_k) \right), \\
\frac{\partial F_{lm}(r)}{\partial t} = N_{lm} \, \sum\limits_{n =
	0}^{\infty} \, 
	\tilde{A}_{nl} \, \frac{d \tilde{\Phi}_{nl}(r)}{dr} \,
	\sum\limits_k \, m_k 
	\frac{\partial}{\partial t} \left( \tilde{\Phi}_{nl}(r_k) \,
	P_{lm}(\cos(\vartheta_k)) \, \sin(m \varphi_k) \right). \\
\intertext{With:}
\begin{split}
\frac{\partial}{\partial t} \, \tilde{\Phi}_{nl}(r_k) &=
	\frac{\partial r_k}{\partial t} \, \tilde{\Phi}_{nl}(r_k) \\
	&\quad \biggl[ \frac{l}{r_k} - \frac{r_k^{\frac{1}{\alpha}}}{r_k} \,
	\frac{2 l + 1}{1 + r_k^{\frac{1}{\alpha}}} + \frac{4
	r_k^{\frac{1}{\alpha}}}{r_k} \, \frac{\alpha \, (2 l + 1) +
	\frac{1}{2}}{\alpha \, (1 + r_k^{\frac{1}{\alpha}})^2} \,
	\frac{C_{n - 1}^{(\omega +
	1)}(\xi_k)}{C_{n}^{(\omega)}(\xi_k)} \biggr], 
\end{split} \\
\frac{\partial}{\partial t} \, P_{lm}(\cos(\vartheta_k)) = -
	\frac{\partial \vartheta_k}{\partial t} \, \sin(\vartheta_k)
	\, \frac{\partial P_{lm}(\cos(\vartheta_k))}{\partial
	\cos(\vartheta_k)}, \\
\frac{\partial}{\partial t} \, \cos(m \varphi_k) = - m
	\frac{\partial \varphi_k}{\partial t} \, \sin(m \varphi_k), \\
\frac{\partial}{\partial t} \, \sin(m \varphi_k) = m
	\frac{\partial \varphi_k}{\partial t} \, \cos(m \varphi_k). 
\end{gather}

in the coefficient computation section the standard leap frog
integrator provided by \citet{Hernquist:92} is
extended by two additional variables, which are computed by using the
recursion relations in section \ref{kap:Rekursion}.

\section{Orbit dependency of the force derivative}
\label{kap:GradPhiDotOrb}

In order to account for the change of force due to the particle orbit
one has to calculate the nine partial derivatives in equation
(\ref{eq:SCFadotorb}). These nine derivatives will now be listed. In
order to save some space the second derivative of
$\tilde{\Phi}_{nl}(r)$ is given first:
\begin{equation}
\begin{split}
\frac{\partial^2}{\partial r^2} \tilde{\Phi}_{nl}(r) &=
	\tilde{\Phi}_{nl}(r) \, \biggl[ \biggl( \frac{l}{r} -
	\frac{r^{\frac{1}{\alpha}}}{r} \, \frac{(2 l + 1)}{(1 +
	r^{\frac{1}{\alpha}})} \biggr)^2 - \frac{l}{r^2} \\
	&\quad + \frac{(2 l + 1)}{(1 + r^{\frac{1}{\alpha}})^2} \,
	\frac{r^{\frac{1}{\alpha}}}{\alpha r^2} \, (\alpha - 1 +
	\alpha r^{\frac{1}{\alpha}}) \\
	&\quad + \biggl[ \, 8 \, \biggl(
	\frac{r^{\frac{1}{\alpha}}}{r} \, \frac{\omega}{\alpha (1 +
	r^{\frac{1}{\alpha}})^2} \biggr) \, \biggl( \frac{l}{r} -
	\frac{r^{\frac{1}{\alpha}}}{r} \, \frac{(2 l + 1)}{(1 +
	r^{\frac{1}{\alpha}})} \biggr) \\
	&\quad + \frac{4 \omega r^{\frac{1}{\alpha}}}{r^2 \alpha^2 (1
	+ r^{\frac{1}{\alpha}})^3} \, (1 - \alpha - (\alpha + 1)
	r^{\frac{1}{\alpha}}) \biggr] \, \frac{C_{n - 1}^{(\omega +
	1)}(\xi)}{C_n^{(\omega)}(\xi)} \\
	&\quad + 16 \, \biggl( \frac{r^{\frac{1}{\alpha}}}{\alpha r}
	\biggr)^2 \, \frac{\omega (\omega + 1)}{(1 +
	r^{\frac{1}{\alpha}})^4} \, \frac{C_{n - 2}^{(\omega +
	2)}(\xi)}{C_n^{(\omega)}(\xi)}
	\label{eq:SCFdtwophinltil}
\end{split}
\end{equation}
% Alte Gleichung (Hernquist-Profil)
%\begin{equation}
%\begin{split}
%\frac{\partial^2}{\partial r^2} \tilde{\Phi}_{nl}(r) & = 
%	\tilde{\Phi}_{nl}(r) \, \biggl[ \biggl( \frac{l}{r} \, - \,
%	\frac{2l - 1}{1 + r} \biggr)^2 \, - \, \frac{l}{r^2} \, + \,
%	\frac{2l + 1}{(1 + r)^2} \, + \\
%	&\quad \biggl( \frac{8 (2l + 3/2)}{(1 + r)^2} \, \biggl(
%	\frac{l}{r} \, - \, \frac{2l - 1}{1 + r} \biggr) \, - \,
%	\frac{8 (2l + 3/2)}{(1 + r)^3} \biggr) \, \frac{C_{n - 1}^{(2 l +
%	5/2)}(\xi)}{C_{n}^{(2 l + 3/2)}(\xi)} \, + \\
%	&\quad \frac{16 (2l + 5/2)(2l + 3/2)}{(1 + r)^4} \, \frac{C_{n
%	- 1}^{(2 l + 7/2)}(\xi)}{C_{n}^{(2 l + 3/2)}(\xi)} \biggr].
%\end{split}
%\end{equation}
The nine derivatives can be implemented as follows:

\subsection{Derivatives with respect to $r$}
The radial derivative for the radial acceleration becomes:
\begin{align}
\frac{\partial a_r}{\partial r} & =  - \sum\limits_{l = 0}^{\infty} \,
	\sum\limits_{m = 0}^{\infty} \, P_{lm}(\cos(\vartheta)) \left[
	G_{lm}(r) \cos(m \varphi) + H_{lm}(r) \sin(m \varphi) \right],
	\\ \intertext{with:}
G_{lm} & = N_{lm} \, \sum\limits_{n = 0}^{\infty} \,
	\tilde{A}_{nl} \frac{\partial^2}{\partial r^2}
	\tilde{\Phi}_{nl}(r) \, \sum\limits_k \, m_k 
	\tilde{\Phi}_{nl}(r_k) \, P_{lm}(\cos(\vartheta_k)) \, \cos(m
	\varphi_k), \\
H_{lm} & = N_{lm} \, \sum\limits_{n = 0}^{\infty} \,
	\tilde{A}_{nl} \frac{\partial^2}{\partial r^2}
	\tilde{\Phi}_{nl}(r) \, \sum\limits_k \, m_k 
	\tilde{\Phi}_{nl}(r_k) \, P_{lm}(\cos(\vartheta_k)) \, \cos(m
	\varphi_k).
\end{align}
The radial deriavtive for the acceleration in $\vartheta$ direction
becomes: 
\begin{equation}
\begin{split}
\frac{\partial a_\vartheta}{\partial r} & = - \sin(\vartheta)
	\sum\limits_{l = 0}^{\infty} \, \sum\limits_{m = 0}^{\infty}
	\, \frac{\partial P_{lm}(\cos(\vartheta))}{\partial
	\cos(\vartheta)} \\ 
	&\quad \times \biggl[
	\biggl(\frac{1}{r^2} \, C_{lm}(r) \, - \, \frac{1}{r} \,
	E_{lm}(r) \biggr) \, \cos(m \varphi) \\
	&\quad + \biggl(\frac{1}{r^2} \,
	D_{lm}(r) \, - \, \frac{1}{r} \, F_{lm}(r) \biggr) \, \sin(m
	\varphi) \biggr]. 
\end{split}
\end{equation}
The radial deriavtive for the acceleration in $\varphi$ direction
becomes: 
\begin{equation}
\begin{split}
\frac{\partial a_\varphi}{\partial r} & = 
	\sum\limits_{l = 0}^{\infty} \, \sum\limits_{m = 0}^{\infty}
	\, \frac{m P_{lm}(\cos(\vartheta))}{\sin(\vartheta)} \\ 
	&\quad \times \biggl[
	\biggl(\frac{1}{r^2} \, D_{lm}(r) \, - \, \frac{1}{r} \,
	F_{lm}(r) \biggr) \, \cos(m \varphi) \\
	&\quad - \biggl(\frac{1}{r^2} \,
	C_{lm}(r) \, - \, \frac{1}{r} \, E_{lm}(r) \biggr) \, \sin(m
	\varphi) \biggr]. 
\end{split}
\end{equation}

\subsection{Derivatives with respect to $\vartheta$}
The derivative with respect to $\vartheta$ for the radial acceleration
becomes: 
\begin{equation}
\frac{\partial a_r}{\partial \vartheta} = \sin(\vartheta) \,
	\sum\limits_{l = 0}^{\infty} \, \sum\limits_{m = 0}^{\infty}
	\, \frac{\partial P_{lm}(\cos(\vartheta))}{\partial
	\cos(\theta)} \left[
	E_{lm}(r) \cos(m \varphi) + F_{lm}(r) \sin(m \varphi) \right].
\end{equation}
The derivative with respect to  $\vartheta$ for the acceleration in
$\vartheta$ direction becomes:
\begin{equation}
\begin{split}
\frac{\partial a_\vartheta}{\partial \vartheta} & = \frac{1}{r} \,
	\sum\limits_{l = 0}^{\infty} \, \sum\limits_{m = 0}^{\infty}
	\, \biggl( \cos(\vartheta) \, \frac{\partial
	P_{lm}(\cos(\vartheta))}{\partial \cos(\vartheta)} \, - \,
	\sin^2(\vartheta) \, \frac{\partial^2
	P_{lm}(\cos(\vartheta))}{\partial \cos(\vartheta)^2} \biggr) \\ 
	&\quad \times [ C_{lm}(r) \cos(m \varphi) + D_{lm}(r) \sin(m
	\varphi) ]. 
\end{split}
\end{equation}
The derivative with respect to $\vartheta$ for the acceleration in $\varphi$
direction becomes:
\begin{equation}
\begin{split}
\frac{\partial a_\varphi}{\partial \vartheta} & = \frac{1}{r} \,
	\sum\limits_{l = 0}^{\infty} \, \sum\limits_{m = 0}^{\infty}
	\, m \, \biggl( \frac{\partial
	P_{lm}(\cos(\vartheta))}{\partial \cos(\vartheta)} \, + \,
	\cos(\vartheta) \,
	\frac{P_{lm}(\cos(\vartheta))}{\sin^2(\vartheta)} \biggr) \\ 
	&\quad \times [ D_{lm}(r) \cos(m \varphi) - C_{lm}(r) \sin(m
	\varphi) ]. 
\end{split}
\end{equation}

\subsection{Derivatives with respect to $\varphi$}
The derivative with respect to $\varphi$ for the acceleration in
radial direction becomes:
\begin{equation}
\frac{\partial a_r}{\partial \varphi} = - \sum\limits_{l = 0}^{\infty}
	\, \sum\limits_{m = 0}^{\infty} \, m P_{lm}(\cos(\vartheta))
	\, [ F_{lm}(r) \cos(m \varphi) - E_{lm}(r) \sin(m \varphi) ].
\end{equation}
The derivative with respect to $\varphi$ for the acceleration in
$\vartheta$ direction becomes:
\begin{equation}
\frac{\partial a_\vartheta}{\partial \varphi} = \frac{\sin(\vartheta)}{r} \,
	\sum\limits_{l = 0}^{\infty} \, \sum\limits_{m = 0}^{\infty}
	\, m \, \frac{\partial P_{lm}(\cos(\vartheta))}{\partial
	\cos(\vartheta)} \,
	\, [ D_{lm}(r) \cos(m \varphi) - C_{lm}(r) \sin(m \varphi) ].
\end{equation}
The derivative with respect to $\varphi$ for the acceleration in
$\varphi$ direction becomes:
\begin{equation}
\frac{\partial a_\varphi}{\partial \varphi} = \frac{1}{r} \,
	\sum\limits_{l = 0}^{\infty} \, \sum\limits_{m = 0}^{\infty}
	\, m^2 \, \frac{P_{lm}(\cos(\vartheta))}{\sin(\vartheta)} \, [
	C_{lm}(r) \cos(m \varphi) + D_{lm}(r) \sin(m \varphi) ].
\end{equation}

\end{document}